\documentclass[twocolumn]{aa}
\usepackage[comma,authoryear]{natbib}
\usepackage[dvips]{graphicx}
\usepackage[english]{babel}
\usepackage[latin1]{inputenc}
\usepackage{latexsym}
\usepackage{rotating}
\usepackage{amssymb,amsmath}

\renewcommand{\bf}{}
\newcommand{\dt}[1]{\frac{\partial  #1}{\partial t}}
\newcommand{\ddpart}[2]{\frac{\partial^2 #1}{\partial #2^2}}
\newcommand{\dpart}[2]{\frac{\partial #1}{\partial #2}}
\newcommand{\ephi}{\vect{e}_{\varphi}}
\newcommand{\er}{\vect{e}_r}
\newcommand{\etheta}{\vect{e}_{\theta}}
\newcommand{\ex}{\vect{e}_x}
\newcommand{\ey}{\vect{e}_y}
\newcommand{\ez}{\vect{e}_z}

\newcommand{\grad}{\vect{\nabla}}
\newcommand{\lapl}{\Delta}
\newcommand{\rot}{\vect{\nabla}\times}
\newcommand{\vect}{\vec}
\newcommand{\Em}{E_{\mathrm{m}}}
\newcommand{\Km}{K_{\mathrm{m}}}
\newcommand{\M}{\mathcal{M}}
\newcommand{\e}{\mathrm{e}}
\newcommand{\E}{\varepsilon}
\renewcommand{\v}{v}
\renewcommand{\S}{\mathcal{S}}
\renewcommand{\.}{\cdot}
\renewcommand{\O}{\mathcal{O}}
\renewcommand{\bar}{\underline}
\renewcommand{\div}{\vect{\nabla}\.}
\renewcommand{\d}{\mathrm{d}}

\renewcommand{\l}{\ell}
\newcommand{\ylm}{Y_{\l}^m}
\newcommand{\Rlm}{\vect{R}_{\l}^m}
\newcommand{\Slm}{\vect{S}_{\l}^m}
\newcommand{\Tlm}{\vect{T}_{\l}^m}
\newcommand{\alm}{a_m^{\l}}
\newcommand{\blm}{b_m^{\l}}
\newcommand{\clm}{c_m^{\l}}
\newcommand{\ulm}{u_m^{\l}}
\newcommand{\vlm}{\v_m^{\l}}
\newcommand{\wlm}{w_m^{\l}}

\newcommand{\alo}{a_0^{\l}}
\newcommand{\blo}{b_0^{\l}}
\newcommand{\clo}{c_0^{\l}}
\newcommand{\ulo}{u_0^{\l}}
\newcommand{\vlo}{\v_0^{\l}}
\newcommand{\wlo}{w_0^{\l}}

\begin{document}
\author{D. Reese \and F. Rincon \and M. Rieutord}
\offprints{D. Reese \\ \email{daniel.reese@ast.obs-mip.fr}}
\institute{
Laboratoire d'Astrophysique de Toulouse, 
Observatoire Midi-Pyrénées,
14 avenue \'E. Belin,
31400 Toulouse,
France}
\date{Received March 29, 2004; accepted July 16, 2004}
\title{Oscillations of magnetic stars}
\subtitle{II. Axisymmetric toroidal and non-axisymmetric shear Alfvén modes in
a spherical shell}
\abstract{We carry out numerical and mathematical investigations of shear
Alfvén waves inside of a spherical shell filled with an incompressible
conducting fluid, and bathed in a strong dipolar magnetic field. We focus on
axisymmetric toroidal and non-axisymmetric modes, in continuation of a previous
work by \hbox{Rincon \&} \hbox{Rieutord (2003, A\&A, 398, 663)}.  {\bf
Analytical expressions are obtained for toroidal eigenmodes and their
corresponding frequencies at low diffusivities. These oscillations behave like
magnetic shear layers, in which the magnetic poles play a key role, and hence
become singular when diffusivities vanish}.  It is also demonstrated that
non-axisymmetric modes are split into two categories, namely poloidal or
toroidal types, following similar asymptotic behaviours as their axisymmetric
counterparts when the diffusivities become arbitrarily small.

}
\maketitle
\section{Introduction}
\addcontentsline{toc}{chapter}{Introduction}

Numerous astrophysical systems exhibit a pulsating behaviour that can
be significantly influenced by the Lorentz force when a strong magnetic
field is present. This may for instance be the case in neutron stars and
magnetic white dwarfs \citep{lou95}. Planetary cores, which are known
to sustain a strong dynamo \citep{stevenson83}, are also likely to fall
into this category.

One of the most exciting examples of couplings between pulsation and
magnetism is given by the seismological activity of roAp stars. This class
of stars, discovered by \cite{1978IBVS}, exhibits several kG (almost)
dipolar magnetic fields and luminosity variations on periods ranging
from 5 to 15 min. These oscillations seem to be well approximated by
a single spherical harmonic $\l = 1$ lined up with the magnetic axis,
suggesting a strong mixing between high order $p$-modes and Alfv\'enic
type oscillations.

Many different models have been developed to obtain a satisfying picture
of the asteroseismology of these stars. Following theoretical
work by \citet{biront,roberts} and \citet{campbell}, \citet{Dziembowski}
have studied acoustic star models enveloped by a layer in which magnetic
effects become dominant. Using a boundary layer approximation, they
came up with an outer boundary condition which was then applied for the
calculation of adiabatic acoustic modes.  This model has undergone a lot
of refinements in order to take into account new physical processes. The
latest version, proposed by \citet{ObliqueModelRevisited}, incorporates
the centrifugal force (a non-axisymmetric effect, since the rotational
and magnetic axis are often tilted in roAp stars) and suggests that
the axis of the modes is not necessarily lined up with the magnetic
axis. In spite of these  improvements, there
are still non-negligible discrepancies between the magnetically shifted
eigenfrequencies computed from these models and the observed ones, showing
that the precise coupling mechanism occurring in the surface layers is
likely to be more complex.  An important point is that a single $\l$ value is
sometimes assumed to be sufficient to describe the oscillations. This
may not be the case, owing to the dipolar structure of the permanent
magnetic field which induces a coupling between spherical harmonics thus
producing a whole spectrum of $\l$'s (e.g. Rincon \& Rieutord 2003).

Motivated by the observation that chemical peculiarities are observed near the
magnetic poles of roAp stars, \citet{excitation} have tried to determine what
precise physical phenomena were occurring in the polar and equatorial regions. 
Since the magnetic field is almost horizontal near the equator and  vertical
near the poles, convection is certainly inhibited in the latter region,
allowing the diffusion of different chemical elements  (an excess of helium is
observed on the polar caps). Oscillations triggered by a $\kappa$-mechanism may
therefore preferentially be observed in this region. This approach stresses the
importance of a global description of the eigenmodes in such stars.

The work presented here originates partly from the preceding remarks. It
continues the study described in Rincon \& Rieutord 2003 (hereafter referred to
as Paper~I) and should be viewed as a preliminary step to obtain global models
of the more complex magneto-gravito-acoustic oscillations. We aim at understanding some
of the underlying physical mechanisms potentially involved in roAp stars or
similar astrophysical objects, by using an approach that rigorously treats the
couplings induced by the geometry of the magnetic field. Our highly simplified
model consists of a non-rotating spherical shell of incompressible magnetised
fluid bathed in a dipolar magnetic field, with small magnetic and kinematic
diffusivities. As a consequence of incompressibility, the modes we compute are
shear Alfv\'en waves.  In Paper~I, a spherical harmonic decomposition of the
linearised MHD equations was obtained and results regarding poloidal
axisymmetric modes were presented (a short reminder of the classification of
modes is given later on in the paper). It was shown that the least-damped modes
are near the magnetic poles and exhibit internal shear layers which can
potentially play a role in mode selection. In the present study, we focus on
the other types of shear Alfv\'enic oscillations in spherical shells, namely
axisymmetric toroidal and non-axisymmetric modes, and characterise numerically
and mathematically their geometry, eigenspectrum, and some by-products like
boundary layers.

In Sect.~2, we recall the basic physical ingredients of the model and shortly
describe our numerical strategy.  Sect.~\ref{sect:tor} is devoted to the
phenomenological description and analytical study of axisymmetric toroidal
modes.  Sect.~\ref{sect:NA} covers non-axisymmetric modes and their resemblance
to their axisymmetric counterparts. Finally, Sect.~5 and~6 conclude the
paper. Note that most of the mathematical details involved in
Sect.~\ref{sect:tor} are given in App.~\ref{appendix:latitudinal} and
App.~\ref{appendix:model} to preserve the clarity of the manuscript.

{\bf \section{Description of the model}} 
\subsection{Basic equations}
We first give a brief description of the model that is used in
calculating toroidal and non-axisymmetric modes and which was already used in
Paper~I.  More details on the basic equations and their expansion onto the
spherical harmonic base are given in Paper~I.

The ``star'' we work with is a spherical shell of incompressible plasma
of density $\rho_o$, with a radius $R$ and an aspect ratio $\eta$. In all
the numerical examples and figures, we use $\eta = 0.35$\footnote{This
is the aspect ratio of the Earth's liquid core or that of the radiative
zone of a 3 M$_\odot$ star.}.  Within this shell is a dipolar magnetic
field generated in a perfectly conducting core:

\begin{equation}
\label{eq:dipole}
   \vect{B} = B_o \. R^3 \left( \frac{\cos \theta}{r^3} \er 
         + \frac{\sin \theta} {2r^3} \etheta \right).
\end{equation}
We use the following underlined dimensionless variables:
$$ \vect{r} = R\bar{\vect{r}}, \quad \vect{\v} = V_A \bar{\vect{\v}}, \quad
\vect{B} = B_o \bar{\vect{B}}, \quad t = \frac{R}{V_A}\bar{t},$$
where $V_A$, the Alfvén velocity, is given by the following expression,
$\mu_o$ being the magnetic permeability:
\begin{equation}
  V_A = \frac{B_o}{\sqrt{\rho_o \mu_o}}.
\end{equation}
We apply the linearised magneto-hydrodynamic equations to the kinetic
$\vect{\v}$ and magnetic $\vect{b}$ perturbations (we drop the underlined
notation).  We assume these perturbations have a time dependence of the form
$\e^{\lambda t}$, where $\lambda = \tau + i \omega$ ($\tau$ is the damping
rate, $\omega$ the pulsation and $i^2=-1$).  This leads to the following set of
equations:
\begin{equation}
\label{eq:system}
\begin{array}{lll}
        \div \vect{\v} &=& 0, \\
        \div \vect{b} &=& 0,  \\
        \lambda \rot \vect{\v}  &=&  \rot \left( (\rot \vect{b}) \times \vect{B}
                \right) + E \rot \lapl \vect{\v}, \\
        \lambda \vect{b} &=& \rot \left(\vect{\v} \times 
                \vect{B} \right) + \Em \lapl \vect {b}.
\end{array}
\end{equation}
where $\vect{B}$ denotes the permanent dipolar magnetic field.  The parameters
$E$ and $\Em$ are non-dimensional forms of the kinematic and magnetic
diffusivities, respectively, and are given by the following expressions:
\begin{equation}
 E = \frac{\nu}{RV_A}, \quad \Em = \frac{1}{\sigma_o \mu_o R V_A},
\end{equation}
where $\sigma_o$ is the conductivity of the fluid. In the case of roAp stars,
these parameters take on the following typical values :
$$ \Em \sim 10^{-8}, \qquad E \sim 10^{-13}. $$

\subsection{Boundary conditions}
The boundary conditions used in our model are as follows.  For the velocity, we
use stress-free, non-penetrative, conditions to minimise the importance of
boundary layers.  This leads to the following conditions, valid for $r = \eta$
and $r = 1$:
\begin{equation}
        \dpart{}{r} \left( \frac{\v_{\theta}}{r} \right) = 0,  \qquad
        \dpart{}{r} \left( \frac{\v_{\varphi}}{r} \right) = 0, \qquad
        \v_r = 0.
\end{equation}
As for the magnetic field, different conditions apply for the inner and outer
boundaries. On the interior, the perturbation to the electric field is
perpendicular to the conducting core, and the perturbation to the magnetic field is
tangent.  This gives the following three equations:
\begin{equation}
\label{condition:conducteur}
\begin{array}{rcl}
         b_r &=& 0, \\
        \noalign{\smallskip}
        \displaystyle
        \frac{\Em}{r} \dpart{}{r}(rb_{\theta}) 
                &=& -\v_{\theta} B_r, \\ 
        \noalign{\smallskip}
        \displaystyle
        \frac{\Em}{r} \dpart{}{r}(rb_{\varphi})  
                &=& -\v_{\varphi}B_r.
\end{array}
\end{equation}
When the other boundary conditions are taken into account, and when
$\lambda$ is different from zero, these equations are not independent and only
correspond to two conditions.  On the outer boundary and beyond, the magnetic
field is continuous and potential, since there are no currents in empty space. 
By using the continuity of this field, and the idea that it does not diverge
toward infinity, it is possible to define two boundary conditions, which are
more easily expressed in spherical harmonics (see Eq.~(\ref{eq:bc})).

Eq.~(\ref{eq:system}), together with these boundary conditions, defines a
generalised eigenvalue problem, where $\lambda$ is the eigenvalue and
$(\vect{\v}, \vect{b})$ is the eigenvector which can be computed numerically.

\subsection{{\bf Harmonic projection}}
In order to solve Eq.~(\ref{eq:system}), the fields $\vect{\v}$ and $\vect{b}$
are projected onto the harmonic base:
\begin{equation}
\label{eq:harmonic_decomposition}
\begin{array}{lcl}
  \vect{\v} &=& \displaystyle \sum_{\l=0}^{\infty} 
    \sum_{m=-\l}^{\l} \ulm \Rlm + \vlm \Slm + \wlm \Tlm, \\
  \vect{b} &=& \displaystyle \sum_{\l=0}^{\infty}
    \sum_{m=-\l}^{\l} \alm \Rlm + \blm \Slm + \clm \Tlm,
\end{array}
\end{equation}
in which $\Rlm$, $\Slm,$ and $\Tlm$ are the normalised spherical harmonics:
$$ \Rlm = \ylm \er, \quad \Slm = r \grad \ylm, \quad \Tlm = r \rot \Rlm. $$
The full harmonic decomposition of Eq.~(\ref{eq:system}) is given in App.~A
of Paper~I. {\bf The outer boundary condition on the magnetic field reads}:
\begin{equation}
\label{eq:bc}
  \begin{array}{l}
  \displaystyle \dpart{a_m^{\l}}{r} + \frac{(\l+2)a_m^{\l}}{r} = 0, \\
  \noalign{\smallskip}
  \displaystyle c_m^{\l} = 0.
  \end{array}
\end{equation}

\subsection{Classification and symmetries}

The various eigenmodes fit naturally into different categories. Firstly, as was
already shown in Paper~I, there is no coupling between different $m$'s.  Hence,
eigenmodes will be made up of only one $m$.  This leads to two types of modes:
axisymmetric oscillations $(m=0)$ and non-axisymmetric ones $(m \neq 0)$. 
Secondly, within the axisymmetric category, it is possible to distinguish
between poloidal modes and toroidal ones, {\bf as the corresponding equations
fully decouple for $m=0$ (see App. A of Paper~I)}.  Poloidal modes are made up
of $\ulo$, $\vlo$, $\alo$, and $\blo$ functions which correspond to $\er$ and
$\etheta$ components.  Toroidal modes are made up of $\wlo$ and $\clo$
functions and are in the $\ephi$ direction.

A certain number of symmetries are present in the physical system and lead to a
few simplifications.  As was pointed out in Paper~I, a parity can be defined
for eigenmodes.  However, there was a slight confusion as to the parity of
toroidal components (see Eq.~(19) of Paper~I) since a mode is called even when
the velocity perturbation is symmetric with respect to the equator and the
magnetic perturbation antisymmetric, and a mode is odd in the reverse
situation.  The corrected form for toroidal eigenvectors reads:
\begin{equation}
\begin{array}{lcll}
(\vect{\v},\vect{b}) &=& (w_o^{2k+1},c_o^{2k}) & \mbox{even,} \\
(\vect{\v},\vect{b}) &=& (w_o^{2k},c_o^{2k+1}) & \mbox{odd.}
\end{array}
\end{equation}

Non-axisymmetric modes also have a global parity.  In fact it is possible to
anticipate this result without the use of spherical harmonics.  If we define
$\M$ as being the operator that gives the mirror image of a vector field with
respect to the equator\footnote{In spherical coordinates, $\M$ is defined as
follows: $\M \vect{V} (r,\theta,\varphi) = V_r(r, \pi - \theta, \varphi) \er -
V_{\theta}(r, \pi - \theta, \varphi) \etheta + V_{\varphi} (r, \pi- \theta,
\varphi) \ephi$.}, and if $(\vect{\v},\vect{b},\lambda)$ is a solution, then
$(\M\vect{\v},-\M\vect{b},\lambda)$ is also a solution.  By combining these two
solutions, it is possible to extract an even part and an odd part from the
original solution. Eigenvectors will take on the following form:
\begin{equation}
\begin{array}{lcll}
(\vect{\v},\vect{b})&=&(u_m^{m+2k},w_m^{m+2k+1},a_m^{m+2k+1},c_m^{m+2k})&\mbox{even,} \\
(\vect{\v},\vect{b})&=&(u_m^{m+2k+1},w_m^{m+2k},a_m^{m+2k},c_m^{m+2k+1})&\mbox{odd.}
\end{array}
\end{equation}
This symmetry enables us to work with half as many components for a given
resolution, which is advantageous from the numerical point of view.  {\bf 
Useful information about the eigenspectra can be deduced from the symmetry of
$\vect{B}$ with respect to any meridian}.  For a given solution
$(\vect{\v},\vect{b}, \lambda)$ it is possible to create a second solution $(\S
\vect{\v},\S \vect{b}, \lambda)$, where $\S$ is the operator\footnote{In
spherical coordinates, $\S$ is defined as follows: $\S \vect{V}
(r,\theta,\varphi) = V_r(r, \theta,-\varphi)\er + V_{\theta}(r,\theta,-\varphi)
\etheta - V_{\varphi} (r,\theta,-\varphi) \ephi$.} that gives the mirror image
of a vector field with respect to the meridian that passes through $\varphi =
0$.  When applied to spherical harmonics, the azimuthal order $m$ changes to
$-m$.  From this we conclude that the eigenspectrum of modes with azimuthal
order $m$ is the same as that of $-m$. {\bf This also leads to the decoupling
of poloidal and toroidal components in the axisymmetric case (the poloidal
modes being symmetric with respect to $\S$ and the toroidal modes
antisymmetric).} By taking into account that $\vect{B}$ is real, we deduce that
the eigenspectrum for $m$ is the conjugate of that for $-m$.  Therefore, each
eigenspectrum is symmetrical with respect to the real axis.  Practically, this
means that we only need to explore eigenvalues for positive azimuthal orders,
and only on the upper half of the complex plane $(\omega \geq 0)$.

{\bf If rotation were taken into account, a number of these symmetries would
break down.  Apart from the situation where the rotation and magnetic axis are
aligned, different $m$'s become coupled and mode parity is lost.
The symmetry with respect to $\S$ is lost even when the rotation axis is lined
up with the magnetic field.}

\subsection{Numerical aspects}

Eigenmodes and eigenvalues are calculated numerically using two different
methods.  The first method, based on a QZ algorithm, gives all the eigenvalues
(for the discretised problem) whereas the second method is an iterative
Arnoldi-Chebyshev algorithm which only computes a selection of
eigenvalues and their corresponding eigenmodes.

In our numerical calculations, we use a simplified version of
Eq.~(\ref{eq:harmonic_decomposition}).  Since different $m$'s are
decoupled, we do not have a summation over the azimuthal order.  Also,
because the divergence of both perturbations is zero, the functions $\vlm$
and $\blm$ can be expressed in terms of $\ulm$ and $\alm$, respectively
(see App.~A of Paper~I).  Furthermore, in the case of axisymmetric
modes, the summation on $\l$ will actually start at $1$ instead of $0$
as a result of our boundary conditions.  In the non-axisymmetric case,
the summation on $\l$ will start at $|m|$ as expected. Finally, the
sum on the spherical harmonics is truncated at $L$.  This leads to the
following formulas:

\begin{equation}
\begin{array}{lcl}
  \vect{\v} &=& \displaystyle \sum_{\l=\l_{min}}^{L} 
    \ulm \Rlm + \vlm \Slm + \wlm \Tlm, \\
  \vect{b} &=& \displaystyle \sum_{\l=\l_{min}}^{L}
    \alm \Rlm + \blm \Slm + \clm \Tlm.
\end{array}
\end{equation}
Each of the functions $\ulm$, $\wlm$,$\alm$ and $\clm$ is written in the
form of a truncated Chebyshev series:
$$ \ulm(r) = \sum_{k=0}^{Nr} \tilde{u}_m^{\l}(k) T_k(r). $$
{\bf Typically, we could reach a spatial resolution of $L=350$, $Nr=930$ for
axisymmetric toroidal modes (when analysing boundary layers) and $L=858$,
$Nr=340$ for non-axisymmetric modes.}

We also define Chebyshev and harmonic spectral {\bf coefficients}:
$$ 
 C_k    = \frac{\displaystyle \max_{\l}   |\tilde{u}_m^{\l}(k)|}
               {\displaystyle \max_{\l,k} |\tilde{u}_m^{\l}(k)|}, \qquad
 C_{\l} = \frac{\displaystyle \max_{k}    |\tilde{u}_m^{\l}(k)|}
               {\displaystyle \max_{\l,k} |\tilde{u}_m^{\l}(k)|}.
$$
They are useful for characterising certain aspects of a mode's structure, as
will be seen in Sect.~\ref{sect:boundary.layer}, as well as the convergence of
the discretization.  

\section{Axisymmetric toroidal modes}
\label{sect:tor}
\subsection{Basic properties}
In this section, we discuss the basic properties of axisymmetric toroidal
modes.  As a reminder, the fields $\vect{\v}$ and $\vect{b}$ are both
orientated in the direction $\ephi$.  From the point of view of spherical
harmonics they are made up of $(\wlo,\clo)$ components.

The equations that govern toroidal modes can be simplified.  It is no longer
necessary to take the curl of the Navier-Stokes equation to remove the
pressure term, because this term vanishes owing to axisymmetry.
This leads to the following, nearly symmetric system:
\begin{equation}
\label{eq:toroidal}
\begin{array}{l*3{cl}}
   \lambda b_{\varphi} &=& (\vect{B} \. \vect{\nabla} ) \v_{\varphi} 
       &-& 3\cos\theta/2r^4\v_\varphi
       &+& \Em \lapl' b_\varphi, \\
   \lambda \v_{\varphi} &=& (\vect{B} \. \vect{\nabla} ) b_{\varphi}
       &+& 3\cos\theta/2r^4b_\varphi
       &+& E \lapl' \v_\varphi.
\end{array}
\end{equation}
{\bf Note that $\div \vect{\v} = 0$, $\div \vect{b} = 0$ and $\lapl' =
\lapl-1/r^2\sin^2\theta$.}

{\bf \subsection{Eigenvalue spectrum}}

\begin{figure}[htb]
\begin{center}
\includegraphics[width=8.8cm]{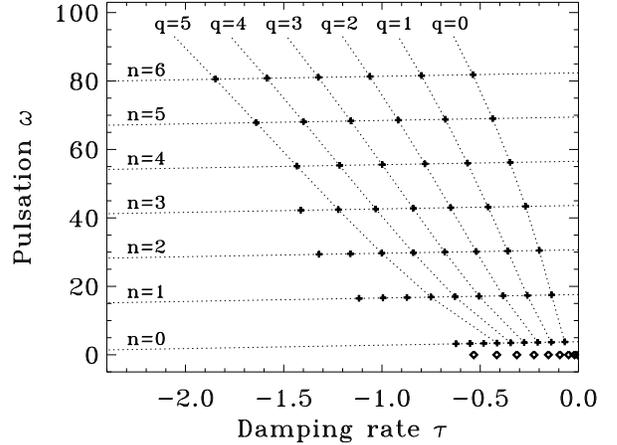}
\end{center}
\caption{Eigenvalue spectrum for odd modes and $E = \Em = 10^{-4}$.  The
plusses ($+$) correspond to oscillatory modes and the diamonds ($\diamond$) to
diffusion modes, which only appear for the odd parity.  The letters $n$ and $q$
quantify the oscillatory mode spectrum.} 
\label{fig:spectre_tor}
\end{figure}

\begin{figure*}
\begin{tabular}{cc}
\includegraphics[width=8.6cm]{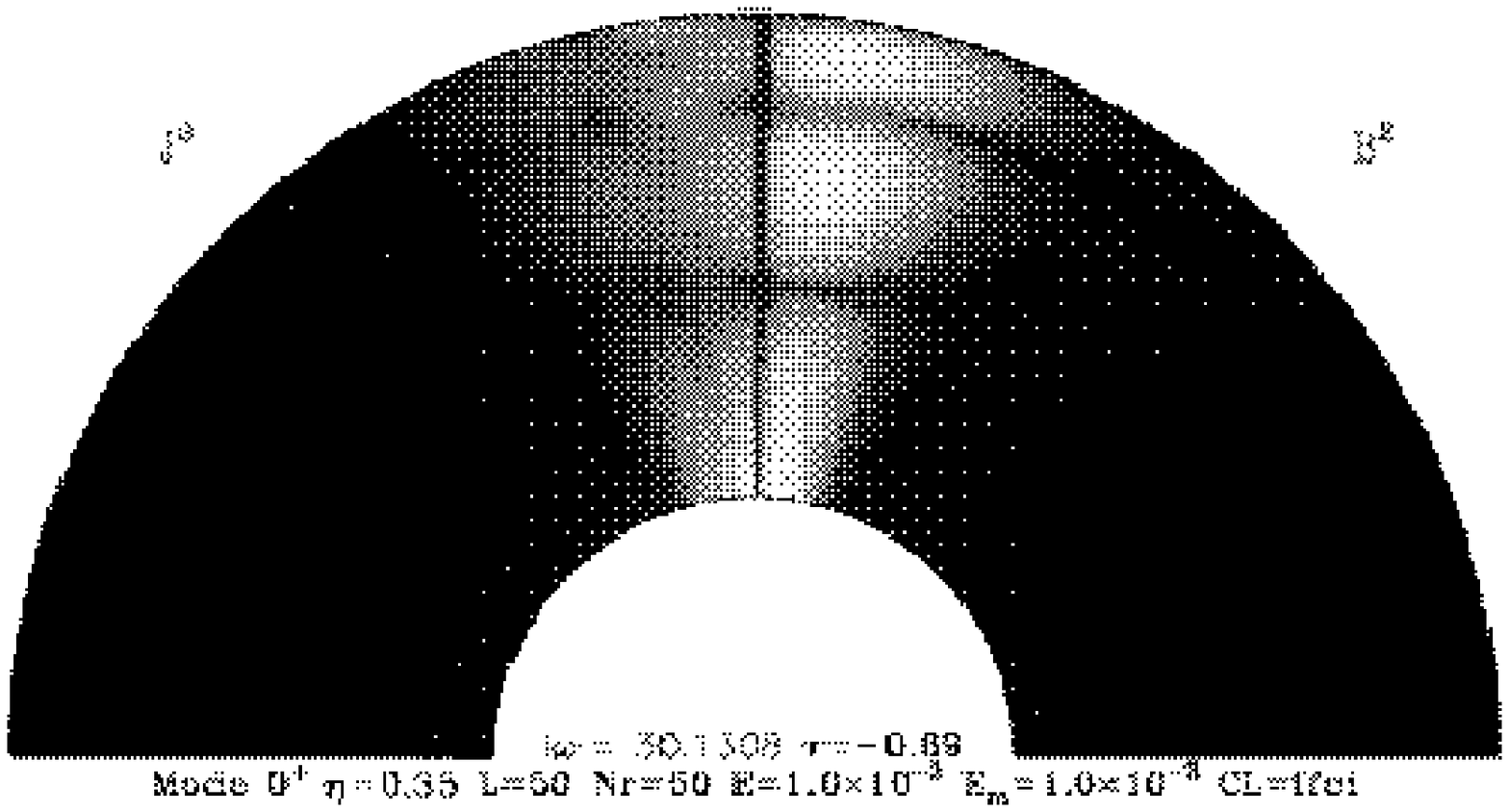} &
\includegraphics[width=8.6cm]{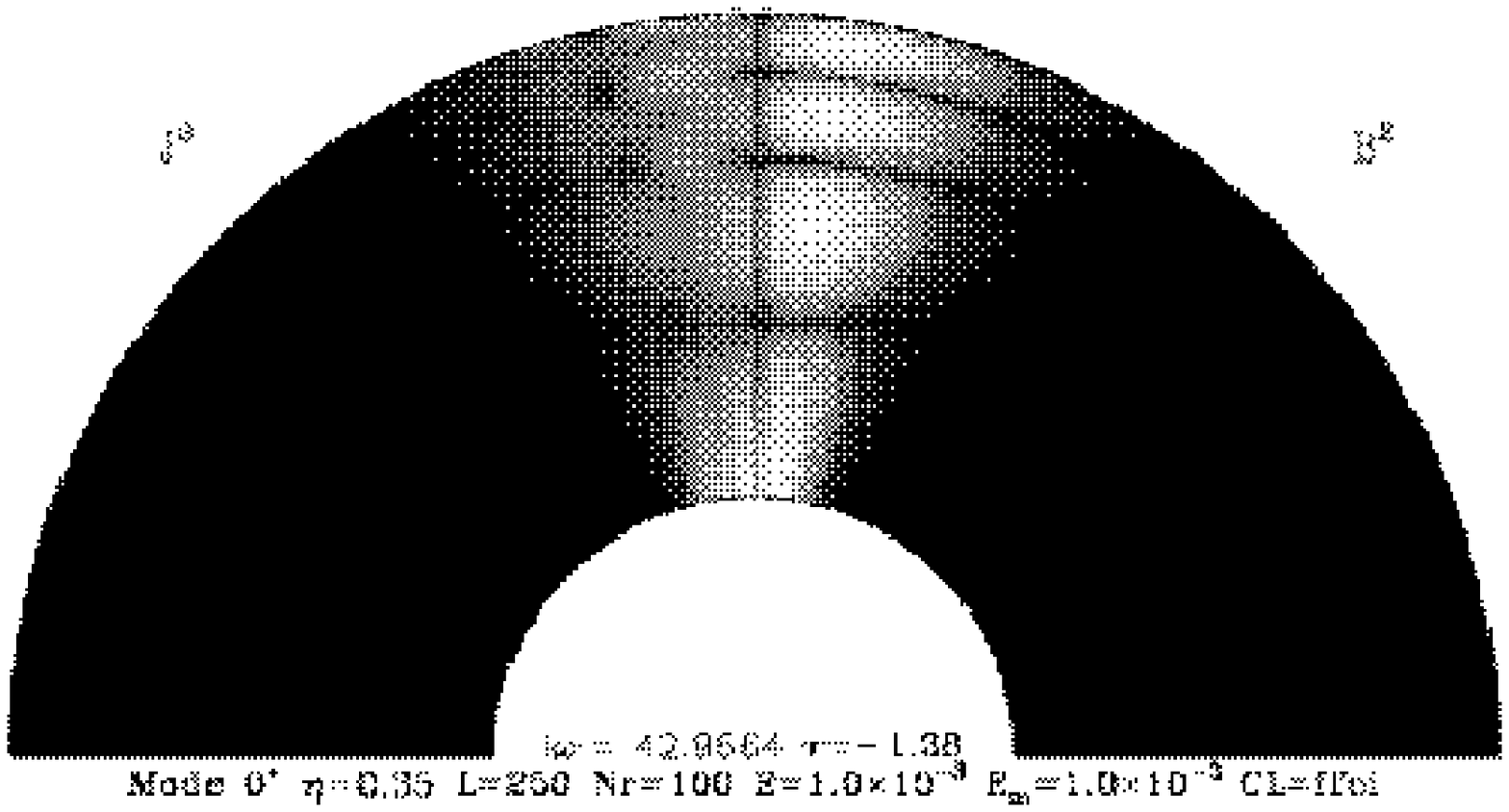} \\
\includegraphics[width=8.6cm]{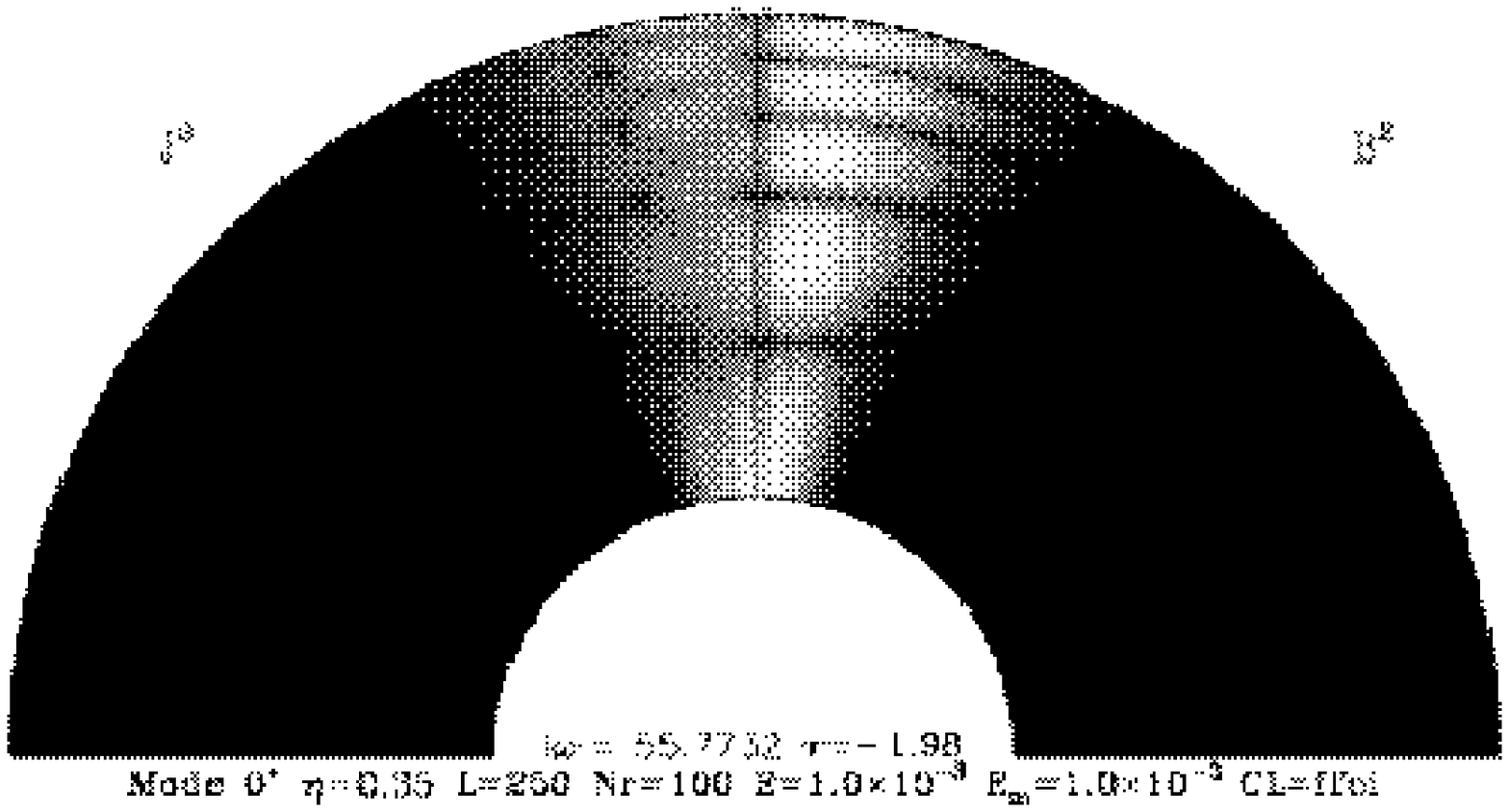} &
\includegraphics[width=8.6cm]{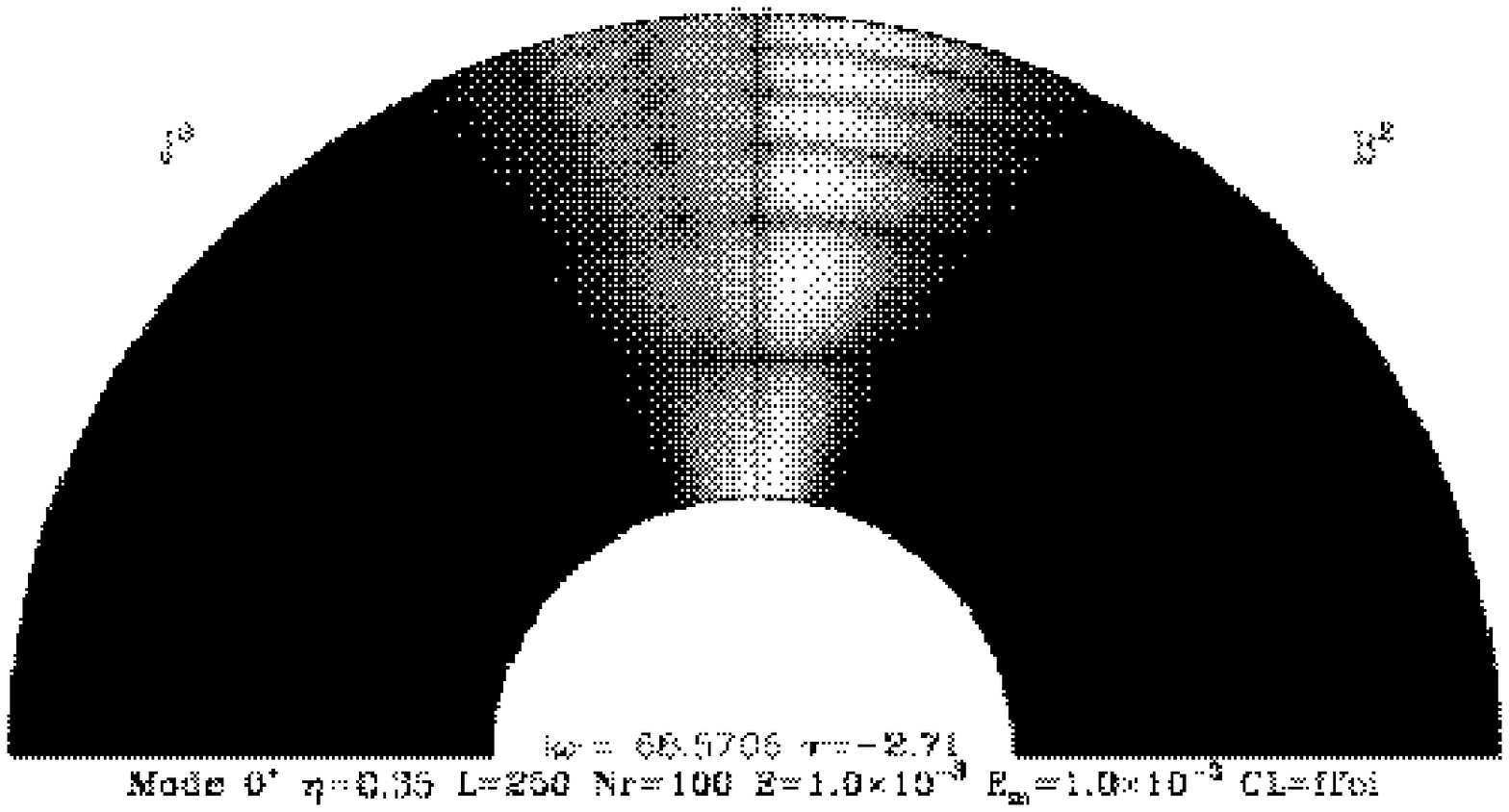} 
\end{tabular}
\caption{Axisymmetric toroidal modes with different vertical structures, for
$E=\Em=10^{-3}$.  The left quadrant in each image corresponds to magnetic
dissipation and the right one to magnetic energy. From left to right, and from
top to bottom, the modes represented here correspond to $(n,q)=(2,0)$, $(3,0)$,
$(4,0)$ and $(5,0)$.  {\bf A logarithmic scale is used in this figure
and most of the other figures in this paper, as it brings out more details on
the modes' structure.}}
\label{fig:vertical.structure}
\end{figure*}

In many ways, the eigenvalue spectrum of toroidal modes is similar to that of
the poloidal modes presented in Paper~I.  The complex eigenvalues are grouped
into ``horizontal'' branches, as can be seen in Fig.~\ref{fig:spectre_tor}. 
These ``eigenbranches'', indexed by $n$, correspond to modes with $n$ nodes in
the radial direction (see Fig.~\ref{fig:vertical.structure}).  Both
perturbations have $n$ nodes, but in different positions.  Depending on
parity, diffusion modes $(\omega = 0,\, \tau \neq 0)$ also appear  but these
modes are of little interest to asteroseismology. In all that
follows, we will only consider oscillatory modes.

A careful look at the eigenvalues along a single branch reveals a regular
structure.  These values are remarkably well lined up, much better than their
poloidal counterparts, and they are equally spaced out.  This suggests the
following empirical law:
\begin{equation}
\label{eq:empirical.eigenbranch}
   \lambda_{n,q} = \alpha_n + q \cdot \beta_n
\end{equation}
where $\alpha_n$ and $\beta_n \, (<0)$ depend on the value of $n$, and $q$ is
given in Fig.~\ref{fig:spectre_tor}.  By looking at the slope of the
eigenbranches, it is possible to determine the ratio between the real and
imaginary parts of $\beta_n$.  In our case, $\mathrm{Re} (\beta_n) \simeq
\mathrm{Im} (\beta_n)$.

\begin{figure}
\begin{tabular}{c}
\includegraphics[width=8.5cm]{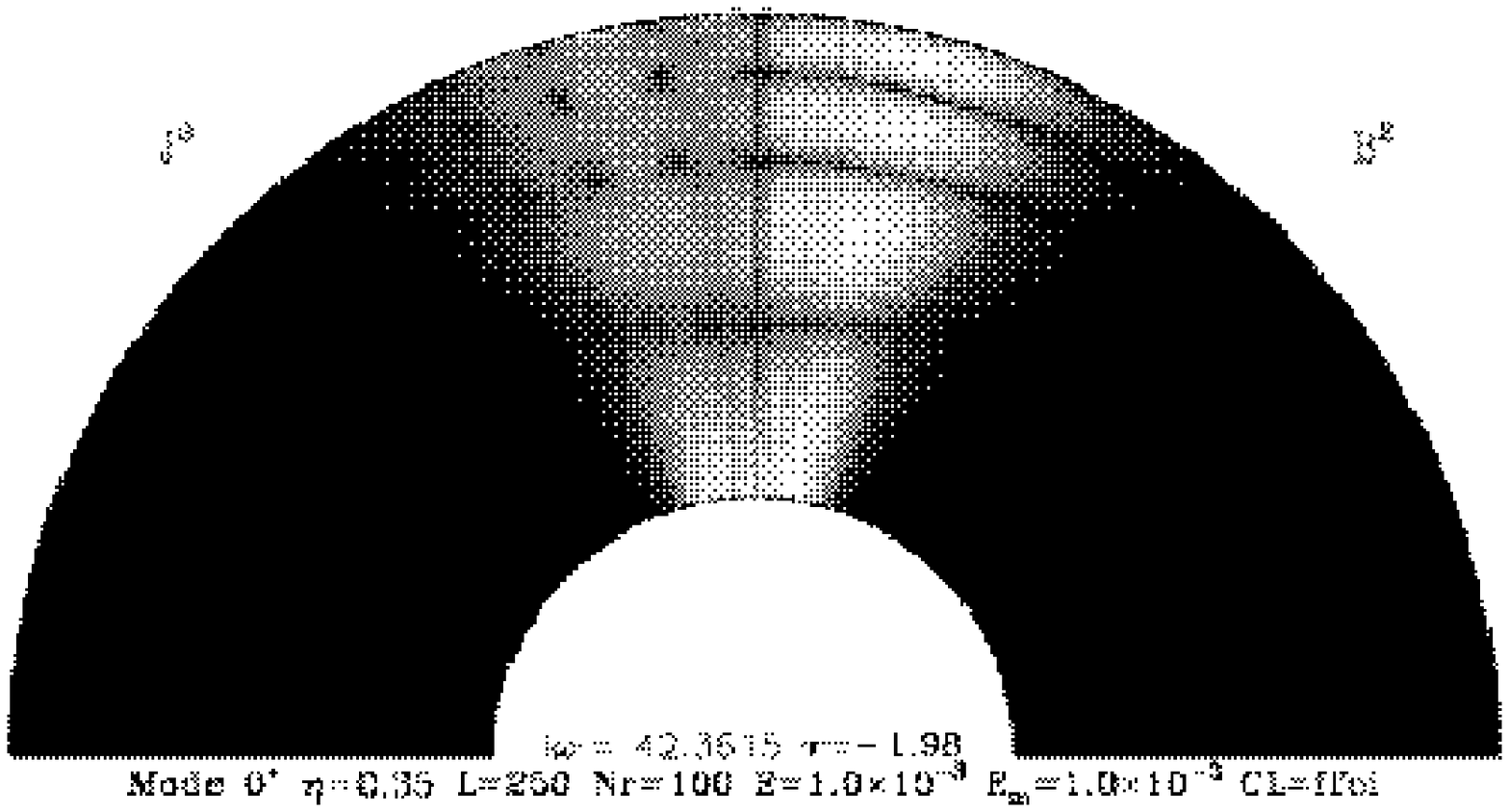} \\
\includegraphics[width=8.5cm]{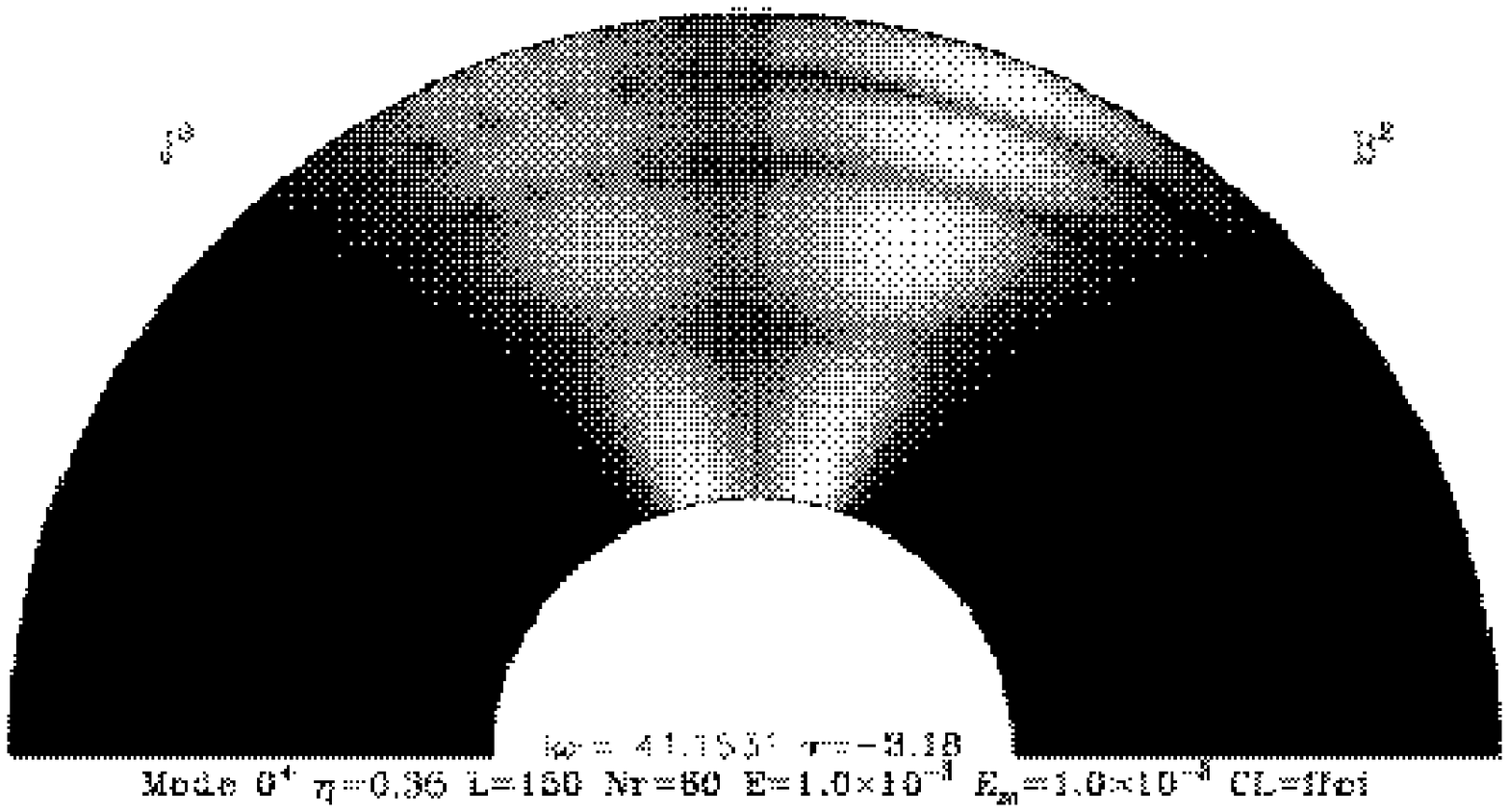} \\
\includegraphics[width=8.5cm]{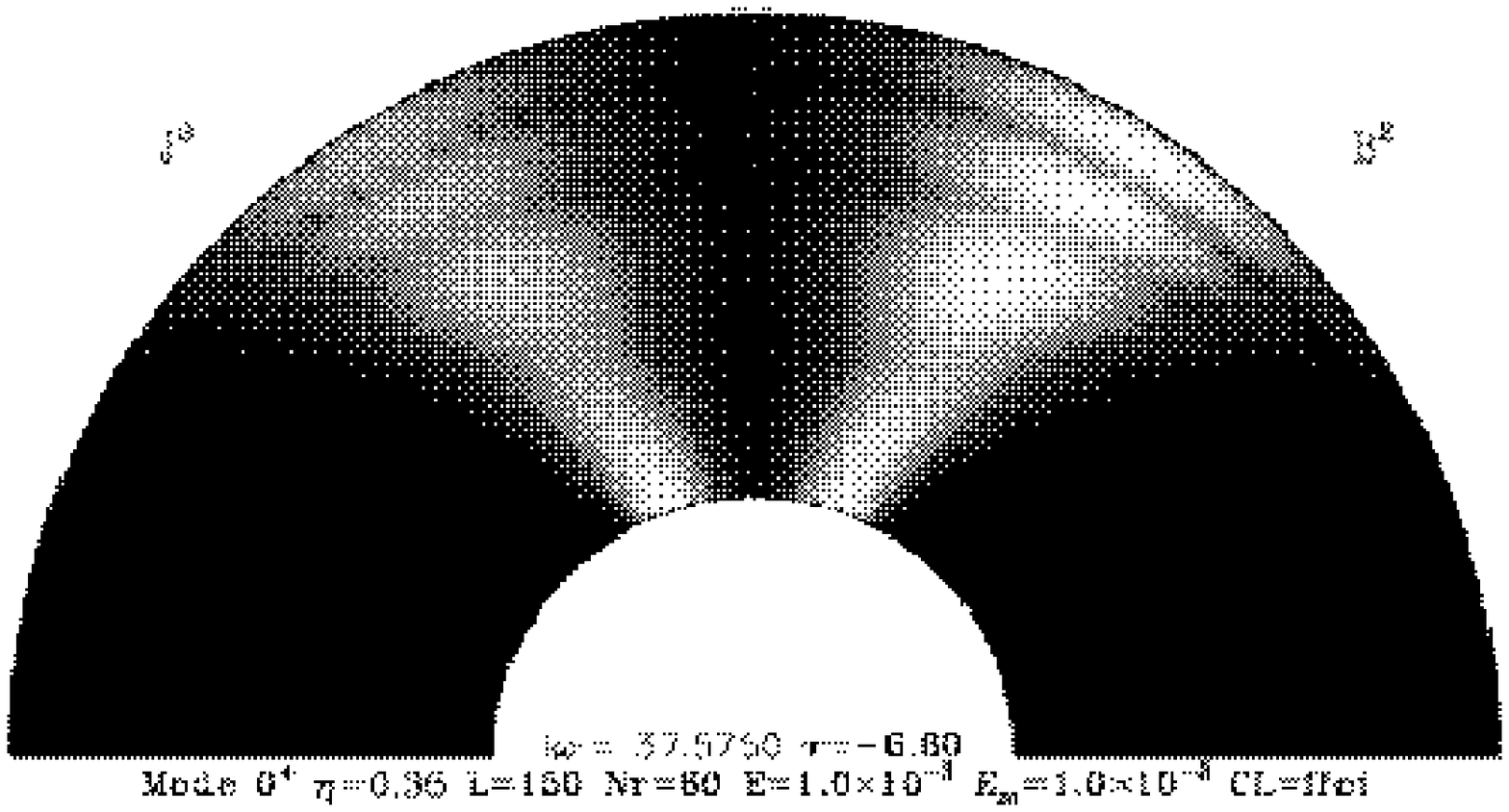} 
\end{tabular}
\caption{Different modes from the $n=3$ eigenbranch. From top to bottom, $q$
takes on the values $1$, $3$ and $9$.}
\label{fig:horizontal.structure}
\end{figure}

We can then look at the corresponding eigenmodes.  When $q$ increases,
two phenomena appear. First of all, for the least-damped modes,
the number of nodes in the horizontal direction seems to increase.
However, these nodes are not all conspicuous.  Secondly, modes that
have a higher damping rate gradually depart from the magnetic poles.
Fig.~\ref{fig:horizontal.structure} illustrates both of these phenomena.
Care must be taken with highly damped {\bf modes:  the spectral coefficients
indicate that the solutions are convergent,} but the corresponding
eigenvalues are highly unstable numerically due to round-off errors.
This limits the number of modes that can be analysed safely.

{\bf \subsection{Mode structure and resonant field lines}}

\begin{figure}
\includegraphics[width=8.5cm,height=4.6cm]{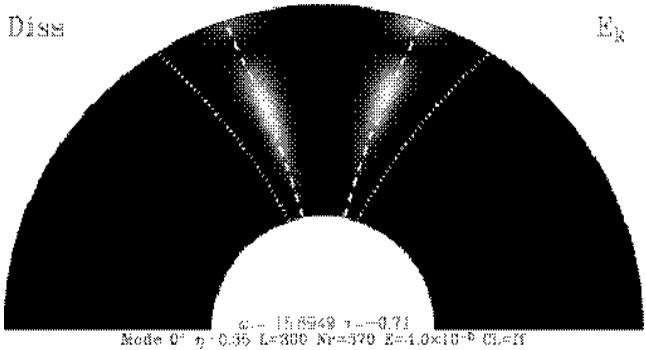}
\caption{{\bf The kinetic energy and dissipation of eigenmode $(n,q) = (1,8)$ with
resonant field lines superimposed.  The diffusivities are $E=\Em=4.10^{-5}$. 
The dotted line corresponds to Eq.~(\ref{eq:alfven.integral}), which is a
WKB-type approximation, and the dashed one to Eq.~(\ref{eq:toroidal.ad}), which
takes the variations of $\vect{B}$ into account.  As opposed to other figures
in this paper, a linear intensity scale is used instead of a logarithmic one. 
This gives a thinner appearance to the mode's structure.}}
\label{fig:resonant.line}
\end{figure}

{\bf The structure of the modes can be understood if one computes
the resonance frequencies of different field lines for
the ideal case $E=\Em=0$.  As each field line can oscillate at its own
frequency,  the resonance frequencies form a continuous spectrum
called the Alfvén continuum.  (For our configuration, there will be a continuum
for each value of $n$).  These eigenvalues are purely imaginary because the
ideal MHD operator which gives the squared eigenvalues $\lambda^2$ is
self-adjoint.

For a first approximation of these frequencies, we can apply the same formula
as was used by \citet{Alfven45}, which is derived from the time it takes for a
wave traveling along a field line to go from one boundary to the other. This
corresponds to a WKB-type approximation, in which the curvature terms have been
neglected:
\begin{equation}
\label{eq:alfven.integral}
	T = \int_{\mathrm{Field~line}} \frac{dl}{\| \vect{V}_{\!\! \mathrm{A}} \|} = 
		\left( n + \frac{1}{2} \right)
                \frac{\pi}{\omega_{\mathrm{f.l.}}},
\end{equation}
where $n$ is the number of nodes along the field line.  A field line will then
be resonant if it satisfies the relation $\omega_{\mathrm{f.l.}} = \omega$
where $\omega$ is the frequency of the mode.   Since the density is constant,
the integral can be calculated analytically and is given by:
\begin{equation}
	\int \! \! \frac{dl}{\| \vect{V}_{\!\! \mathrm{A}} \|}\! = \! \!
	\left\{ \! \!
	\begin{array}{l}
                \displaystyle 
		\frac{2}{\sin^8 \theta_1} {\left[
		\frac{\cos^7 \theta}{7} \! - \! \frac{3 \cos^5 \theta}{5} \! + \!
		\cos^3 \theta \! - \! \cos \theta
                \right]}_{\theta_{\eta}}^{\theta_1}
	\\
                \displaystyle
		\frac{1-\eta^4}{4} \qquad \mathrm{if} 
                                   \quad \theta_{\eta} = \theta_1 = 0,
	\end{array}
	\right.
\label{eq:integrale}
\end{equation}
with $[f(\theta)]_{\theta_{\eta}}^{\theta_1} =f(\theta_1) - f(\theta_{\eta})$.
$\theta_{\eta}$ and $\theta_1$ are the colatitudes of the field line on the
inner and outer boundaries and satisfy the relation \hbox{$\sin^2 \theta_{\eta}
= \eta \sin^2 \theta_1$}.  This formula is the same as Eq.~(26) by
\citet{Kato58}, except for the bounds on the integral. We see that the
oscillation rate decreases for field lines further from the magnetic poles,
hence one can expect eigenmodes with lower eigenfrequencies to be further from
the poles.

Taking the variations of $\vect{B}$ into account, in order to obtain better
results, yields the following equations, which still only apply to individual
field lines:
\begin{equation}
\label{eq:toroidal.ad}
\begin{array}{rcl}
  i\omega_{\mathrm{f.l.}} b  &=& \displaystyle
                 \frac{\cos \theta}{r^3} \frac{\d \v}{\d r}
                -\frac{3\cos \theta}{2 r^4} \v, \\
  \noalign{\smallskip}
  i\omega_{\mathrm{f.l.}} \v &=& \displaystyle
                 \frac{\cos \theta}{r^3} \frac{\d b}{\d r}
                +\frac{3\cos \theta}{2 r^4} b, \\
   \noalign{\smallskip}
   \v(\eta) &=& 0, \\
   \noalign{\smallskip}
   b(1)    &=& 0.
\end{array}
\end{equation}
where $\cos \theta =  \sqrt{1 - r \sin^2 \theta_1}$.  The boundary conditions
are derived from a simple analytical model (see
Eq.~(\ref{eq:effective.conditions})) and were used to establish the $n+1/2$
quantization in Eq.~(\ref{eq:alfven.integral}).  With this approach, the
resonant field line coincides much better with the mode (see
Fig.~\ref{fig:resonant.line}).}

\subsection{Polar eigenspectrum}
\label{section:polar.eigenspectrum}

We will now give a precise calculation {\bf (in the adiabatic case
$(E=\Em=0)$)} of the resonance frequencies corresponding to the field line
along the magnetic axis. {\bf This analysis is motivated by the role of
resonating polar field lines in the asymptotic limit of small diffusivities.
The adiabatic eigenfrequencies will be denoted by the superscript ``$0$''.

These frequencies are solutions of Eq.~(\ref{eq:toroidal.ad}) with $\cos \theta
= 1$.} By combining the two first equations and solving for $\v$, {\bf we obtain}:
\begin{equation}
  r^2 \frac{\d^2 \v}{\d r^2} - 3 r \frac{\d \v}{\d r}  +
  \left( \frac{15}{4} + \omega^2 r^8 \right) \v = 0.
\end{equation}
{\bf (We have dropped the notation $\omega_{\mathrm{f.l.}}$.)
The solution to this equation is (see \citet{Abramowitz})}:
\begin{equation}
   \v =   Ar^2 J_{ \frac{1}{8}} \left( \frac{\omega r^4}{4} \right) +
          Br^2 J_{-\frac{1}{8}} \left( \frac{\omega r^4}{4} \right), \\
\end{equation}
where $A$ and $B$ are two constants and $J_{\nu}$ is the Bessel function of the
first kind of order $\nu$. The corresponding solution for $b$ is:   
\begin{equation}
   b =   iBr^2 J_{ \frac{7}{8}} \left( \frac{\omega r^4}{4} \right) -
         iAr^2 J_{-\frac{7}{8}} \left( \frac{\omega r^4}{4} \right),
\end{equation}
The boundary conditions determine the eigenfrequencies, via the following
relation:
\begin{equation} 
\label{eq:eigenvalues.polar}
   J_{ \frac{1}{8}} \left( \frac{\omega {\eta}^4}{4} \right) 
   J_{ \frac{7}{8}} \left( \frac{\omega}{4} \right)
 + J_{-\frac{1}{8}} \left( \frac{\omega {\eta}^4}{4} \right)
   J_{-\frac{7}{8}} \left( \frac{\omega}{4} \right) = 0.
\end{equation}
We will use the notation $\lambda_n^0 = i \omega_n^0$ to mean the $n$th
eigenvalue calculated with this model.  Asymptotically, for high frequencies,
these values are well approximated by the WKB frequencies found in Paper~I:
$\omega_n^0 \sim \frac{2(2n+1)\pi}{1-\eta^4}$.  The corresponding solutions are
written $b_n^0$ and $\v_n^0$ and are given by the formulas:
\begin{equation}
\label{eq:solution.polar}
\begin{array}{lcl}
   \v^0_n &=& \displaystyle
          r^2 J_{ \frac{7}{8}} \left( \frac{\omega^0_n}{4} \right)
              J_{ \frac{1}{8}} \left( \frac{\omega^0_n r^4}{4} \right) \\
   \noalign{\smallskip}
     & & \displaystyle 
     + \, r^2 J_{-\frac{7}{8}} \left( \frac{\omega^0_n}{4} \right)
              J_{-\frac{1}{8}} \left( \frac{\omega^0_n r^4}{4} \right), \\
   \noalign{\smallskip}
   b^0_n &=& \displaystyle
         ir^2 J_{-\frac{7}{8}} \left( \frac{\omega^0_n}{4} \right)
              J_{ \frac{7}{8}} \left( \frac{\omega^0_n r^4}{4} \right) \\
   \noalign{\smallskip}
     & & \displaystyle
    - \, ir^2 J_{ \frac{7}{8}} \left( \frac{\omega^0_n}{4} \right)
              J_{-\frac{7}{8}} \left( \frac{\omega^0_n r^4}{4} \right).
\end{array}
\end{equation}

{\bf It is worth noting the striking similarity between these expressions and
Eq.~(30) by \citet{Kato58}, which corresponds to the radial variation of
geomagnetic poloidal modes (instead of toroidal ones) calculated for a
dipolar background in which the $\theta$-dependence of $\| \vect{B} \|$ has
been neglected.}

In order to show the physical meaning of these eigenvalues, we took
an eigenspectrum and prolongated the lines formed by the eigenbranches
(see Fig.\ref{fig:spectre_tor}).  The intersections between the lines
and the {\bf imaginary} axis $(\tau = 0)$ correspond to what could be
called ``numerical polar eigenfrequencies''.  The agreement between
these values and $\lambda^0_n$ is rather good, as can be seen in
Tab.~\ref{table:polar.eigenspectrum}.  This leads to the conclusion that
the polar eigenfrequencies are a good indicator of the eigenbranches'
positions.

\begin{table}
  $$ 
      \begin{array}{*{4}c}
         \hline
         \noalign{\smallskip}
         n                   &
         \mbox{Num.}         &  
         \omega^0_n          & 
         \mbox{WKB}         \\
         \noalign{\smallskip}
         \hline
         \noalign{\smallskip}
         0 &  3.8167 &  3.8171 &  6.3789 \\
         1 & 17.6147 & 17.5996 & 19.1367 \\
         2 & 30.6762 & 30.6363 & 31.8945 \\
         3 & 43.6358 & 43.5575 & 44.6524 \\
         4 & 56.5614 & 56.4305 & 57.4102 \\
         5 & 69.4727 & 69.2767 & 70.1680 \\
         6 & 82.3822 & 82.1057 & 82.9258 \\
         7 & 95.2896 & 94.9228 & 95.6836 \\
         \noalign{\smallskip}
         \hline
      \end{array}
  $$ 
   \caption{A comparison between numerical polar eigenfrequencies (Num.),
   adiabatic ones ($\omega^0_n$) and a WKB approximation.  The numerical
   values were obtained using a least square method on the eigenvalues
   corresponding to $E=\Em=10^{-4}$ and correspond to the intersection between
   the eigenbranches and the imaginary axis.}
   \label{table:polar.eigenspectrum}
\end{table}

\subsection{Asymptotic behaviour for small diffusivities}
\label{sec:asymptotic_behaviour}

The next step is to look at the behaviour of eigenmodes and eigenvalues when
the diffusivities $E$ and $\Em$ approach zero.  There are two important reasons
for this.  First of all, realistic astrophysical values of $E$ and $\Em$ are
{\bf very small ($\lesssim 10^{-8}$) and still out of reach for numerical
solvers.}  Being able to extrapolate the behaviour of eigenmodes gives an
educated guess as to what they would actually be for low diffusivities.  A
second reason is that it may be possible this way to refine analytically the
adiabatic eigenvalues $\lambda_n^0$ defined by Eq.~(\ref{eq:eigenvalues.polar})
to obtain a better approximation of the true eigenvalues.

\begin{figure}[htbp]
\includegraphics[width=8.8cm]{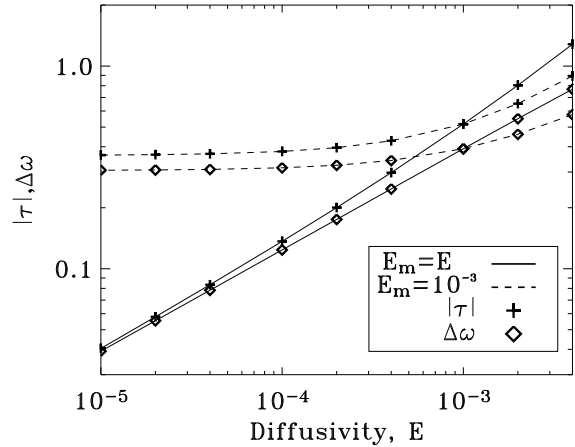} \\
\caption{Behaviour of eigenvalue $(n,q)=(1,0)$ for different values of $E$ and
$\Em$. $\Delta \omega$ corresponds to the difference between the theoretical
polar eigenfrequency, solution to Eq.~(\ref{eq:eigenvalues.polar}), and the
actual frequency of the mode. When $\Em=10^{-3}$, the eigenvalue never attains
$\lambda^0_n$.  When $E=\Em$, the slope of the $\Delta \omega$ line is $1/2$,
meaning that $\Delta \omega$ is proportional to $E^{1/2}$.}
\label{fig:eigenvalue}
\end{figure}

\begin{figure}[htbp]
\includegraphics[width=8.8cm]{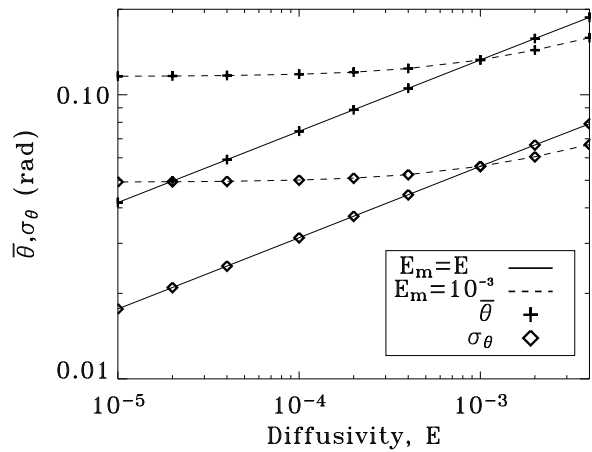}
\caption{Behaviour of an eigenmode for different values of $E$ and $\Em$.
$\overline{\theta}$ is an estimate of the mode's position and $\sigma_{\theta}$
an estimate of the thickness.  In the case $E=\Em$ the position and thickness
seem to be approaching zero; the slope of both lines is $1/4$.}
\label{fig:ecarttype}
\end{figure}

We therefore analysed the behaviour of the mode $n=1$, $q=0$ for different
values of the diffusivities.  Figure~\ref{fig:eigenvalue} and
Fig.~\ref{fig:ecarttype} show how its eigenvalue, its position and its
``thickness'' vary.  In order to calculate the mode's position and thickness,
we took a profile of the magnetic energy $| \vect{b} |^2$ along a meridional
cut with a fixed radius\footnote{The choice of this radius is arbitrary.  If we
pick different radii, we get approximately the same empirical laws (see
Eq.~(\ref{eq:empirical})).} of $0.5$ and a horizontal extent going from
$\theta=0$  to $\theta=\pi/2$.  From this profile, we calculated a mean value
$\overline{\theta}$ and a standard deviation $\sigma_{\theta}$.  These
correspond to estimates of the mode's position and thickness, respectively. 

From these graphs it is possible to deduce some of the physical phenomena that
are taking place.  In the case where $E=\Em$, we observe that the eigenvalue
approaches the $n=1$ polar eigenfrequency given by
Eq.~(\ref{eq:eigenvalues.polar}).   At the same time, the mode seems to get
thinner and closer to the magnetic poles.  We observe from
Fig.~\ref{fig:eigenvalue} and Fig.~\ref{fig:ecarttype} that:
\begin{equation}
\label{eq:empirical}
\begin{array}{lcl}
   \lambda_{1,0} &=& \lambda^0_1 + E^{1/2} (\tau^1 + i \omega^1) + E
                     \tau^{2}, \\
   \overline{\theta} & \propto & E^{1/4}, \\
   \sigma_{\theta} & \propto & E^{1/4},
\end{array}
\end{equation}
in which $\lambda^0_1$ is the polar eigenvalue, and $\tau^{1}$, $\omega^{1}$
and $\tau^{2}$ are constants.  It turns out that $\tau^1 = \omega^1$.  These
laws suggest that toroidal modes become singular when $E=\Em$
approach zero thus revealing their true nature, namely that of magnetic
shear layers.

By identifying Eq.~(\ref{eq:empirical.eigenbranch}) with
Eq.~(\ref{eq:empirical}), we find that:
\begin{equation}
   \alpha_1 = \lambda_1^0 + E^{1/2} (\tau^1 + i \omega^1) + E \tau^2.
\end{equation}
A comparison between $\beta_1$ and $\tau^1+i \omega^1$ permits the
identification:

\begin{equation}
   \beta_1  = E^{1/2} (\tau^1 + i \omega^1).
\label{edemi}
\end{equation}
Therefore, we can cast Eq.~(\ref{eq:empirical.eigenbranch}) into a new form,
which separates the adiabatic contribution from the non-adiabatic ones:
\begin{equation}
\label{eq:empirical.eigenbranch.bis}
   \lambda_{n,q} = \lambda^0_n + E^{1/2} (q+1) (\tau^1_n + i \omega^1_n)
               + E \tau^2_n. \\
\end{equation}
The third term, $\tau^2_n$, is roughly proportional to the square of the
frequency and can therefore become noticeable when $n$ is sufficiently large. 
If the eigenvalues were reduced to the first two terms, then for arbitrary
values of $E$, they would all fall on the lines given by:
\begin{equation}
  \frac{\mathrm{Im}(\lambda)-\omega_n^0}{\omega_n^1} = 
  \frac{\mathrm{Re}(\lambda)}{\tau_n^1}.
\label{eq:eigenlines}
\end{equation}

When $\Em$ is fixed, the situation is different.  The eigenvalue
does not appear to converge toward its corresponding polar eigenvalue.  At the
same time, the mode structure does not become singular, which is consistent
with the behaviour of the eigenvalue.  The same experiment can be repeated, but
with $E$ fixed instead of $\Em$.  The same phenomenon appears.  However, the
limit eigenvalue for $(\Em = 0,E=10^{-3})$ is not the same as for $(\Em =
10^{-3},E=0)$.

\subsection{Asymptotic solutions}
Having separated the adiabatic and non-adiabatic contributions to the
eigenvalue, we will now give a justification of
Eq.~(\ref{eq:empirical.eigenbranch.bis}), based on an analytical model.  We
assume that the diffusivities are of the form $E=K\E$ and $\Em=\Km\E$, with
$\E$ approaching zero.  We find that to zeroth order in $\E$, eigenmodes can be
put in the following form (see App.~\ref{appendix:latitudinal}):
\begin{equation}
\begin{array}{lcl}
  b^{0}(r,\hat{\nu}) &=& b^0_n(r)f(\hat{\nu}), \\
  \v^{0}(r,\hat{\nu}) &=& \v^0_n(r)f(\hat{\nu}),
\end{array}
\end{equation}
where $(b^0_n,\v^0_n)$ are given by Eq.~(\ref{eq:solution.polar}), and
$\hat{\nu} = \E^{-1/4} \sin \theta /\sqrt{r}$ remains constant along
field lines.  We prove in App.~\ref{appendix:latitudinal} that the function
$f$ satisfies the following differential equation (see Eq.~(\ref{eq:latitudinal})):
\begin{equation}
\label{eq:latitudinal.equation}
  \lambda_n^1 \mathcal{C}_1 f =
- \frac{\lambda_n^0 \hat{\nu}^2 \mathcal{C}_2}{2}f
+ \mathcal{C}_3 \left[ \frac{\d^2 f}{\d \hat{\nu}^2} 
                     + \frac{1}{\hat{\nu}}\frac{\d f}{\d \hat{\nu}}
                     - \frac{f}{\hat{\nu}^2} \right],
\end{equation}
{\bf where $\lambda_n^0$ and $\lambda_n^1$ are the two first orders of $\lambda$,
and $\mathcal{C}_1$, $\mathcal{C}_2$ and $\mathcal{C}_3$ are constants given by
the following formulas:}
\begin{equation}
\begin{array}{lcl}
\mathcal{C}_1 &=& \displaystyle
 \int_{\eta}^1 \! r^3 \left( |b^0_n|^2 + |v^0_n|^2 \right) \, dr, \\
\noalign{\smallskip}
\mathcal{C}_2 &=& \displaystyle
 \int_{\eta}^1 \! r^4 \left( |b^0_n|^2 + |v^0_n|^2 \right) \, dr, \\
\noalign{\smallskip}
\mathcal{C}_3 &=& \displaystyle
 \int_{\eta}^1 \! \Km |b^0_n|^2 + K |v^0_n|^2  \, dr.
\end{array}
\end{equation}

The function $f$ vanishes at the pole since $\vect{b}$ and $\vect{\v}$ are
toroidal.  In order to mimic the confinement of eigenmodes near the poles, we
also impose the condition that $f$ approaches zero as $\hat{\nu}$ goes to
infinity. We then look for the eigenfunctions $f$ and eigenvalues $\lambda^{1}_n$
that satisfy Eq.~(\ref{eq:latitudinal.equation}).  The variable $\hat{\nu}$ is
rescaled to $\rho = \hat{\nu} \e^{i\pi /8} \sqrt[4] {\frac{\omega_n^0
\mathcal{C}_2}{2 \mathcal{C}_3}}$ where $\lambda_n^0 = i \omega_n^0$. This
leads to the following equation:

\begin{equation}
\label{eq:diff.longi}
  \frac{\d^2 f}{\d\rho^2}
+ \frac{1}{\rho}\frac{\d f}{\d\rho}
- \left(
         \frac{1}{\rho^2}
       + \rho^2 
       + \frac{\lambda_n^{1} \e^{-i\pi/4} \mathcal{C}_1} {\sqrt{\omega_n^0
       \mathcal{C}_2 \mathcal{C}_3/2}}
  \right) f = 0,
\end{equation}

\begin{figure*}[htb]
\begin{tabular}{cc}
\includegraphics[width=8.8cm]{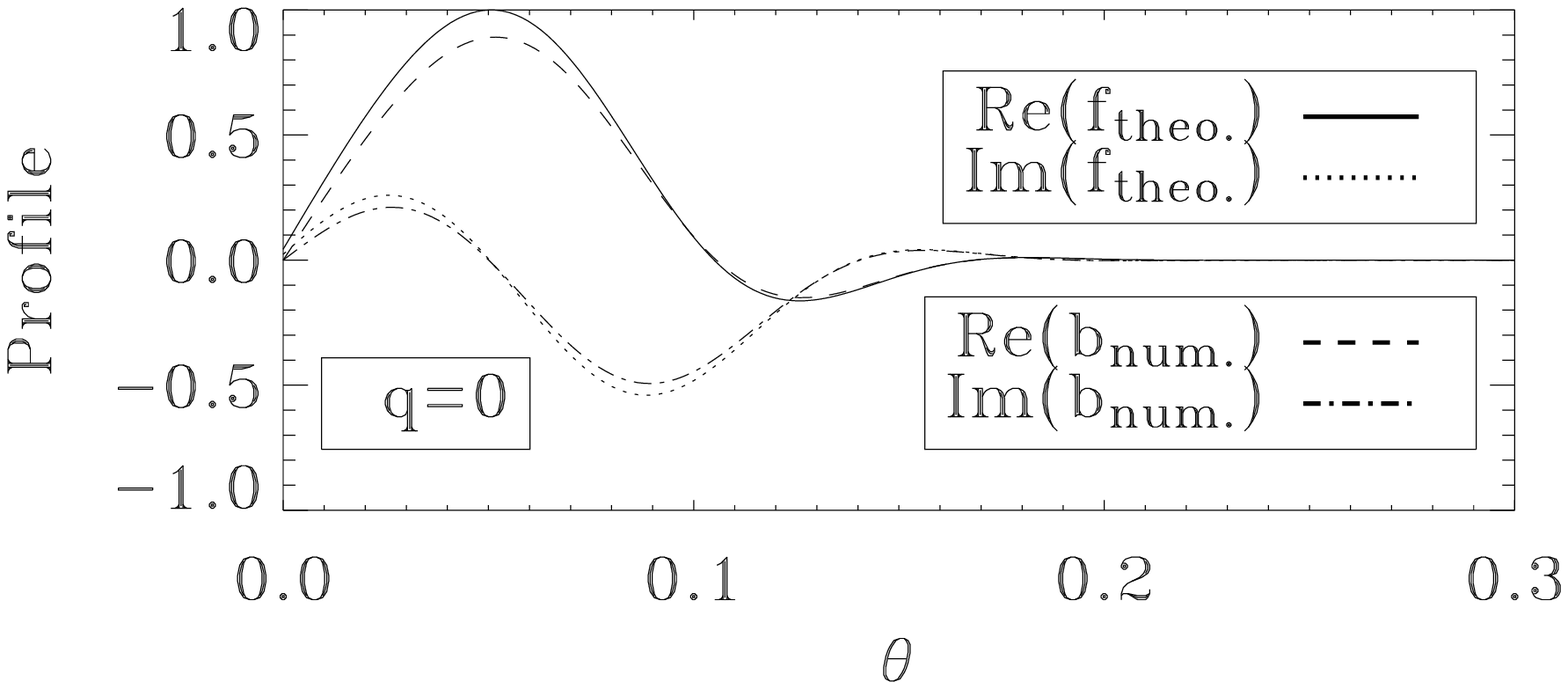} &
\includegraphics[width=8.8cm]{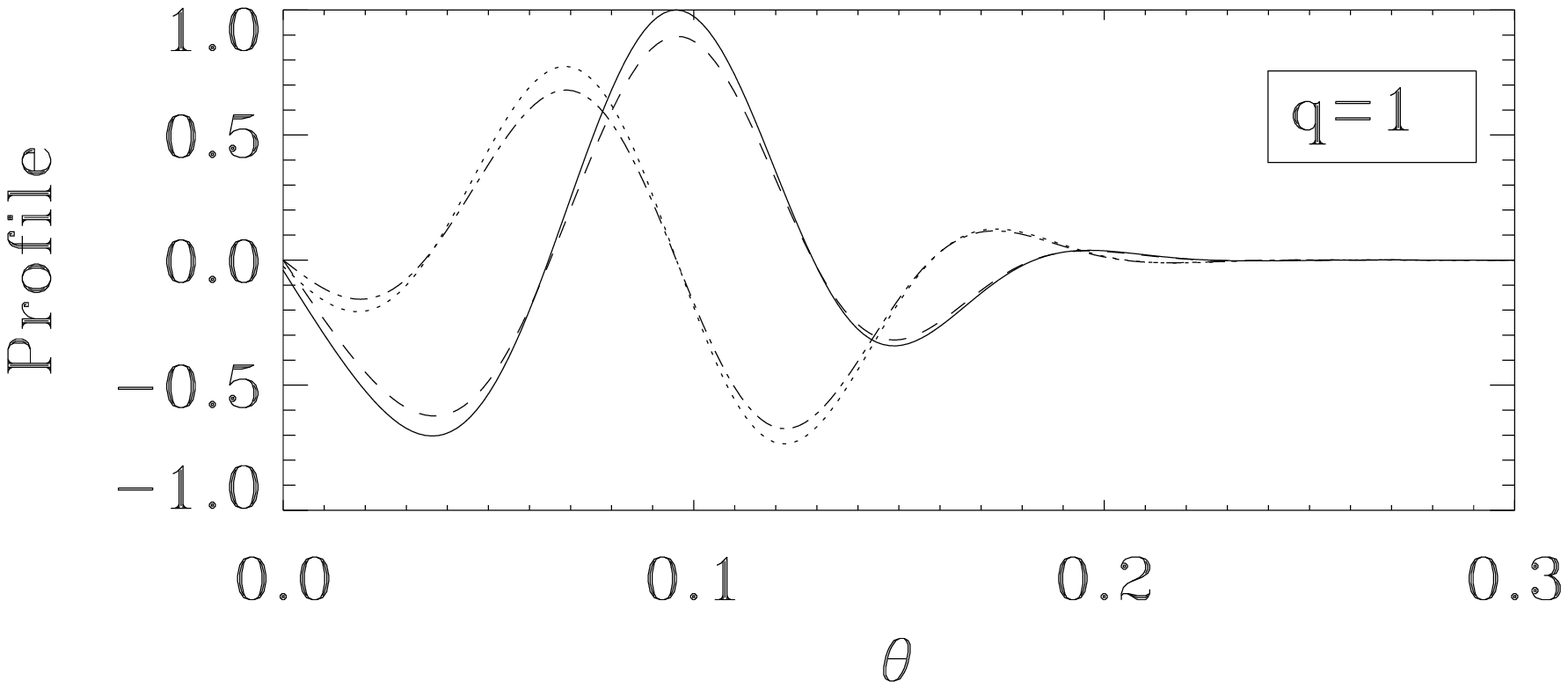} \\
\includegraphics[width=8.8cm]{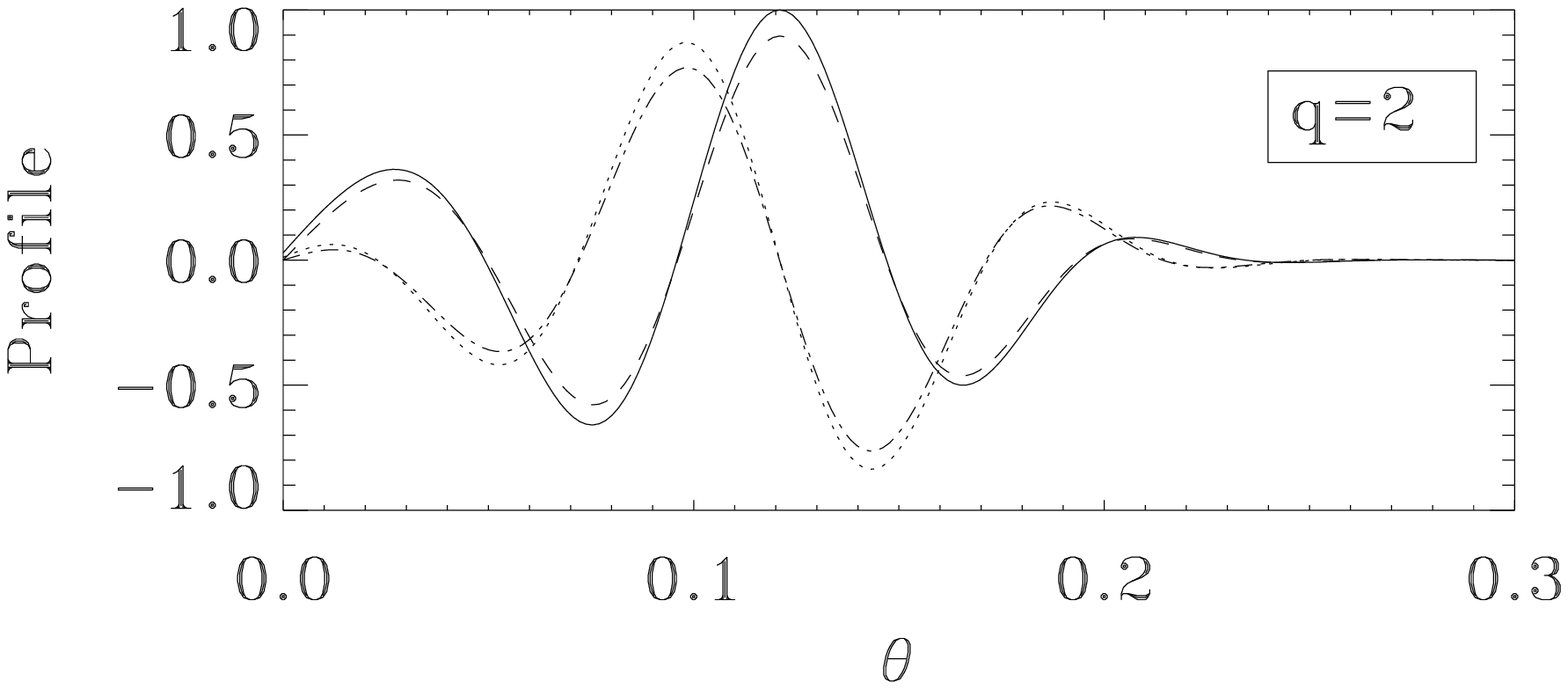} &
\includegraphics[width=8.8cm]{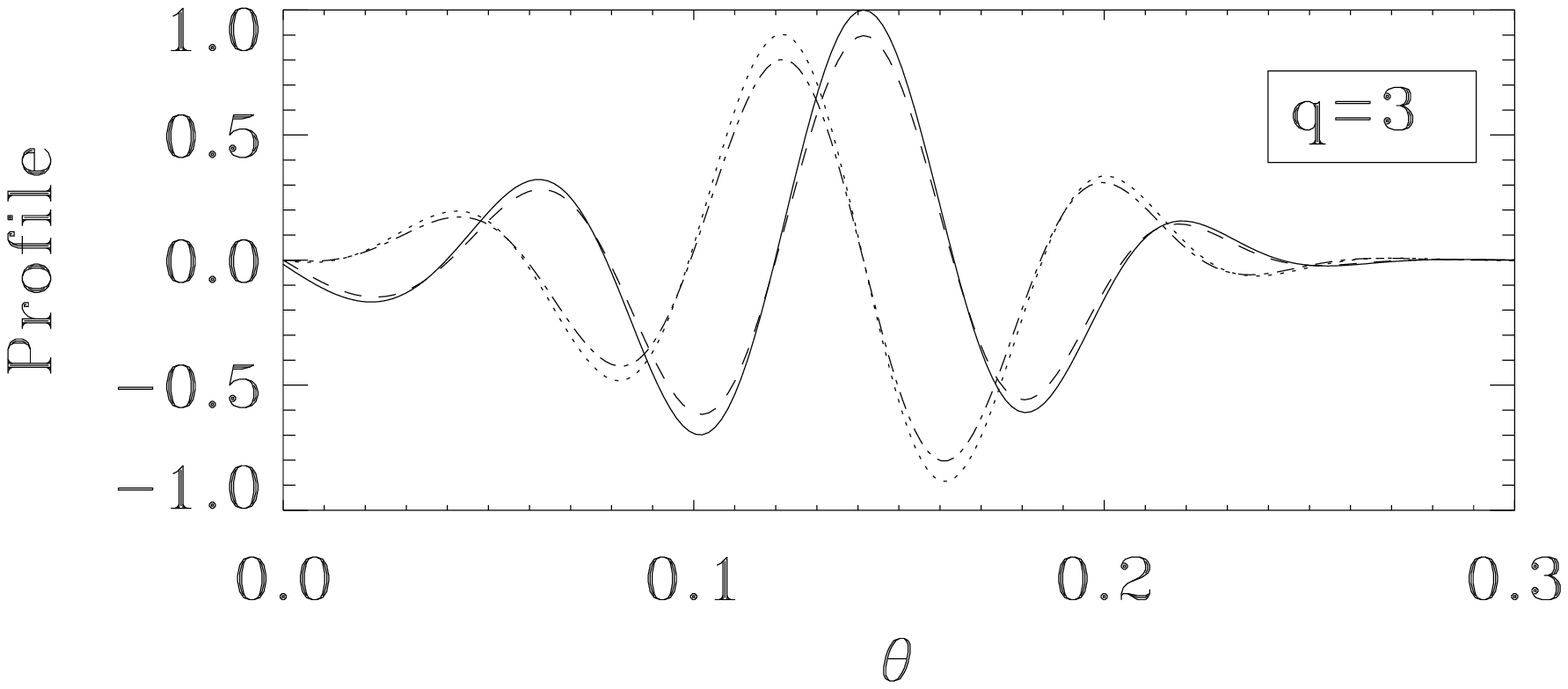}
\end{tabular}
\caption{Comparison between different numerical profiles of the magnetic field
and theoretical ones.  These profiles are calculated along a meridional cut at
a radius $r \simeq 0.5$, and $E=\Em=4.10^{-5}$.}
\label{fig:profiles}
\end{figure*}

A general solution $f$ of Eq.~(\ref{eq:diff.longi}) can be expressed via a
linear combination of the following functions:
\begin{equation}
\begin{array}{lcl}
   s_1(\rho) &=& \rho \, \e^{-\rho^2/2} M \left( a,2,\rho^2 \right), \\
   s_2(\rho) &=& \rho \, \e^{-\rho^2/2} U \left( a,2,\rho^2 \right), \\
   a &=& \displaystyle 1 + \frac{\lambda_n^{1} \e^{-i\pi/4} \mathcal{C}_1}
                {\sqrt{8 \omega_n^0 \mathcal{C}_2 \mathcal{C}_3}},
\end{array}
\end{equation}
where $M$ and $U$ are confluent hypergeometric functions, solutions to Kummer's
equation (see \citet{Abramowitz}, $M$ is known as Kummer's function).  When $a$
is not a negative integer, it can be shown that $s_1$ diverges when $\rho$ goes
to infinity and $s_2$ diverges when $\rho$ approaches $0$.  Hence, the only way
to satisfy our set of boundary conditions is to assume {\bf that $a$ is a}
negative integer, namely $a=-q$.  In that case, the functions $M$ and $U$ both
become proportional to the generalised Laguerre polynomials (see
\citet{Abramowitz}):
\begin{equation}
M(-q,2,x) = \frac{L_q^{(1)}(x)}{q+1} = \frac{x^{-1}\e^x}{(q+1)!}
\frac{\d^q}{\d x^q} \left( x^{1+q}\e^{-x} \right).
\end{equation}
In summary, we obtain the following expressions for toroidal axisymmetric
eigenvalues:
\begin{equation}
\label{eq:tor.eigenvalues}
\begin{array}{lcl}
\lambda_{n,q}     &=& \lambda^{0}_n + \E^{1/2} \lambda^{1}_{n,q} + \O(\E), \\
\noalign{\smallskip}
\lambda^{0}_n     &=& \displaystyle i\omega^0_n \simeq i \frac{2(2n+1)\pi} 
                                                        {1-\eta^4}, \\
\lambda^{1}_{n,q} &=& \displaystyle -2(1+q)(1+i) \frac{\sqrt{\omega^0_n
                      \mathcal{C}_2 \mathcal{C}_3}}{\mathcal{C}_1}.
\end{array}
\end{equation}
The corresponding modes are given by:
\begin{equation}
\label{eq:tor.eigenmodes}
\begin{array}{lcl}
b                 &=& b^0 + \O(\E^{1/2}), \\
\noalign{\smallskip}
\v                &=& \v^0 + \O(\E^{1/2}), \\
\noalign{\smallskip}
b^{0}(r,\theta)   &=& b^0_n(r) \rho \, \e^{-\rho^2/2} L_q^{(1)}(\rho^2), \\
\noalign{\smallskip}
\v^{0}(r,\theta)  &=& \v^0_n(r) \rho \, \e^{-\rho^2/2} L_q^{(1)}(\rho^2),
\end{array}
\end{equation}
in which the variable $\rho$ is given by the following expression:
\begin{equation}
\rho = \E^{-1/4} \e^{i\pi/8} \left\{ \frac{\omega^0_n \mathcal{C}_2}
       {2\mathcal{C}_3} \right\}^{1/4} \frac{\sin \theta}{\sqrt{r}}.
\end{equation}

The form of Eq.~(\ref{eq:tor.eigenvalues}) corresponds
perfectly to that of Eq.~(\ref{eq:empirical.eigenbranch.bis}). A
quantitative comparison reveals the accuracy of these formulas (see
Tab.~\ref{table:first.order.comparison}). It is particularly interesting
to compare the profiles along meridional cuts with these theoretical
predictions. Figure~\ref{fig:profiles} shows such a comparison,
for different values of $q$, a comparison which turns out to be
excellent. Naturally, the analytical formulas are more accurate when
both diffusivities take on small values.

\begin{table}
  $$ 
      \begin{array}{*{4}c}
         \hline
         \noalign{\smallskip}
         n                              &
         \mbox{Num.}(\E^{1/2} \tau^1)   &  
         \mbox{Num.}(\E^{1/2} \omega^1) &  
         \mbox{Theo.}         \\
         \noalign{\smallskip}
         \hline
         \noalign{\smallskip}
         0 & -0.0690 & -0.0685 & -0.0688 \\
         1 & -0.1224 & -0.1237 & -0.1239 \\
         2 & -0.1601 & -0.1608 & -0.1608 \\
         3 & -0.1907 & -0.1910 & -0.1910 \\
         4 & -0.2171 & -0.2173 & -0.2172 \\
         5 & -0.2406 & -0.2406 & -0.2407 \\
         6 & -0.2621 & -0.2624 & -0.2621 \\
         7 & -0.2822 & -0.2819 & -0.2820 \\
         \noalign{\smallskip}
         \hline
      \end{array}
  $$ 
   \caption{Comparison between numerical and theoretical first order
   eigenvalues ($\E^{1/2} \lambda^1$).  The two sets of numerical values
   (columns 2 and 3) are based on the numerical eigenspectrum given in
   Fig.~\ref{fig:spectre_tor} and are calculated using a least square method. 
   These values should be very close to each other.  The last column is based
   on Eq.~(\ref{eq:tor.eigenvalues}).}
   \label{table:first.order.comparison}
\end{table}

\subsection{Boundary layers}
\label{sect:boundary.layer}

To complete our study of axisymmetric toroidal eigenmodes, we also
present some results regarding the existence of boundary layers.
In our case, we expected the presence of Hartmann layers, which are very
thin. Typically their non-dimensional thickness is given by $B_r^{-1}
\sqrt{E\Em}$ in which $B_r$ is non dimensional \cite[e.g.][]{CoucheHartmann}.
This can be penalising for numerical calculations as a high resolution
is needed to properly resolve them.

Two different methods were used to study boundary layers. The first
approach consists in coming up with a highly simplified analytical model.
In this model, developed in App.~\ref{appendix:model} for clarity reasons,
the permanent magnetic field is constant and vertical, and the fluid
domain is enclosed between two parallel planes.  This model leads to the
conclusion that there should be a Hartmann layer on the lower boundary
and nothing on the upper plane.  This Hartmann layer only produces
a finite discontinuity of the normal gradient of both perturbations
when the diffusivities vanish (see Eq.~(\ref{eq:solutions})).  The model
also justifies \textit{a posteriori} the effective boundary conditions
$v(\eta)=0$ and $b(1)=0$ (see Eq.~(\ref{eq:effective.conditions}))
which were used in the calculation of polar eigenvalues (see
Sect. \ref{section:polar.eigenspectrum}).

The second approach is numerical.  By looking at the Chebyshev spectral
{\bf coefficients}, we can see the signature of Hartmann layers.  Assuming the layer
is described by an exponential variation as proposed by \citet{CoucheHartmann},
it is possible to come up with a corresponding {\bf theoretical prediction}, which
can then be compared with the actual Chebyshev spectral {\bf coefficients} of the
eigenmodes.

\begin{figure}[htbp]
\begin{center}
\includegraphics[width=8.8cm]{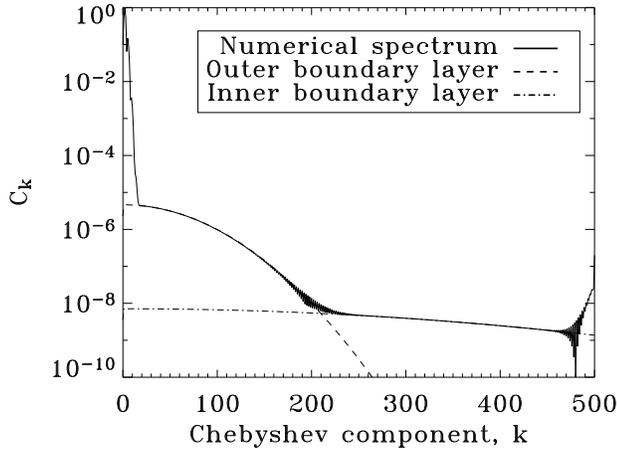}
\end{center}
\caption{Comparison between numerical $(E=\Em=10^{-4})$ and theoretical
spectral {\bf coefficients} which confirms the presence of boundary layers.}
\label{fig:boundary_layer.spectral_decomposition}
\end{figure}

Figure~\ref{fig:boundary_layer.spectral_decomposition} shows such a
comparison.  The values used for $B_r$ are those at the magnetic poles, because
of the mode's localization. In other words, the theoretical inner boundary has
a thickness of $\eta^3 \sqrt{E\Em}$ and the outer one a thickness of $\sqrt
{E\Em}$ (see Eq.~(\ref{eq:dipole})).  The theoretical boundary layers are
multiplied by a constant so as to match the numerical ones.  From this figure,
we clearly deduce the presence of two boundary layers, one on the inside and
one on the outside, unlike what was predicted by the analytical model.  This is
so probably because in the simplified model, the two outer boundary conditions
become redundant when $E$ and $\Em$ approach zero, whereas they do not in the
spherical setup.

For larger diffusivities, the agreement between the numerical and theoretical
results is not as good.  This is not too surprising as the eigenmodes are in
general further from the poles.  At smaller diffusivities, the radial
resolution rapidly becomes insufficient for good comparisons.  The theoretical
spectral {\bf coefficients} for the inner boundary {\bf diminish very little} for the
first several hundred Chebyshev components.

A final comment can be made about the numerical approach.  By looking at the
relative amplitude of the layer on the boundary, it is possible to determine
what type of discontinuity it produces when both diffusivities approach zero. 
In our case, the amplitudes for both the inner and outer layers, obtained by
comparing the numerical and theoretical spectral {\bf coefficients}, were
proportional to $E$ in the case where $E=\Em$, which suggests that there will
be a discontinuity on the gradient of the perturbations when both diffusivities
approach zero.  This is the same behaviour as the single boundary layer in the
analytical model.

\section{Non-axisymmetric modes} 
\label{sect:NA}
We now focus on non-axisymmetric modes $(m \neq 0)$.  From the numerical point
of view, they are twice as demanding as their axisymmetric counterparts for a
given resolution: the $(\ulm,\alm)$ components are coupled to the $(\wlm,\clm)$
components, which gives birth to eigenvectors twice as large.

\subsection{Eigenvalue spectrum}

\begin{figure}[htb]
\begin{tabular}{c}
\includegraphics[width=8.5cm]{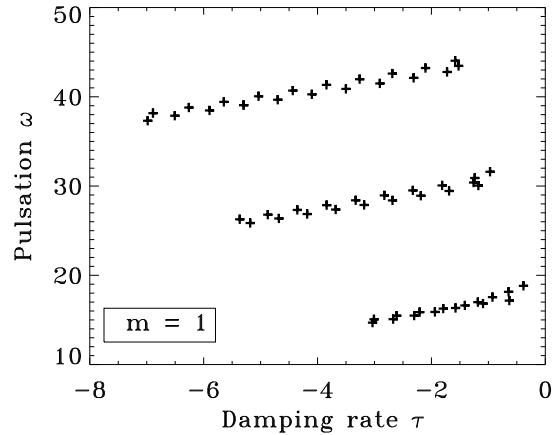} \\
\includegraphics[width=8.5cm]{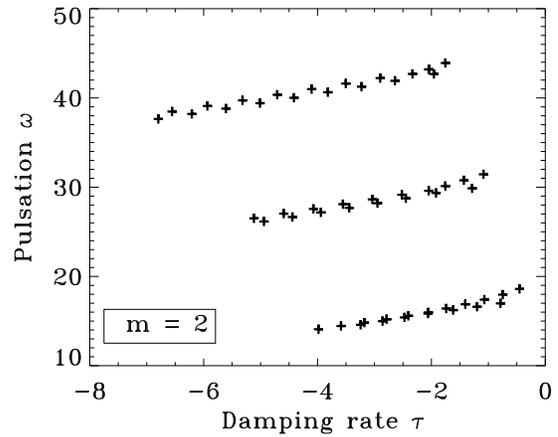} \\
\includegraphics[width=8.5cm]{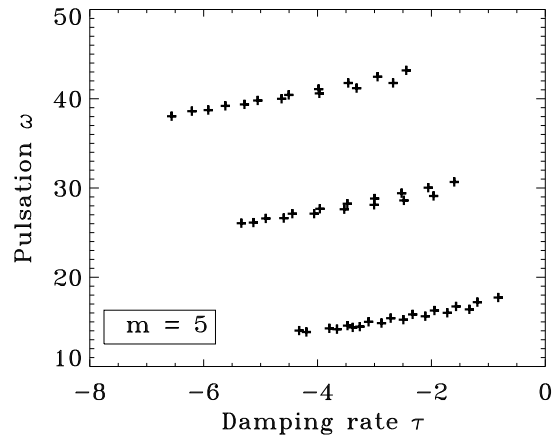} 
\end{tabular}
\caption{Several non-axisymmetric eigenvalue spectra.  The diffusivities $E$
and $\Em$ both take on the value $10^{-3}$.  These branches overlap with the
axisymmetric ones.}
\label{fig:NAspectrum}
\end{figure}

As was the case for poloidal and toroidal eigenspectra, non-axisymmetric
eigenvalues are located along ``horizontal'' branches.  However, where there
used to be one branch in the poloidal or toroidal case, there are now two
branches next to each other for each $m \neq 0$ (see
Fig.~\ref{fig:NAspectrum}).  This is perhaps not surprising as non-axisymmetric
modes contain poloidal and toroidal components at the same time.

\begin{figure}[htb]
\includegraphics[width=8.8cm]{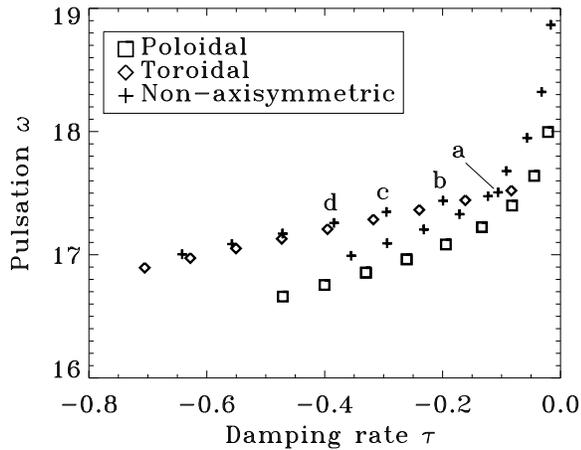}
\caption{Comparison between axisymmetric poloidal, axisymmetric toroidal and
non-axisymmetric $(m=1)$ eigenvalues.  The diffusivities $E$ and $\Em$ both
take on the values $4.10^{-5}$.  The lower non-axisymmetric branch
corresponds to poloidal-like modes.  The upper branch contains toroidal-like
modes when $\tau$ is sufficiently negative.  The letters ``a'',``b'',``c'' and
``d'' label non-axisymmetric modes that are represented in
Fig.~\ref{fig:transition}.}
\label{fig:comparison}
\end{figure}

As diffusivities become small, the two branches start to differ.  When compared
with poloidal or toroidal branches, it becomes immediately obvious that a first
group of eigenvalues resembles the toroidal branch and a second group matches its
poloidal counterpart (see Fig.~\ref{fig:comparison}).  

\begin{figure*}[htb]
\begin{tabular}{cc}
Poloidal modes & Toroidal modes \\
\includegraphics[width=8.6cm]{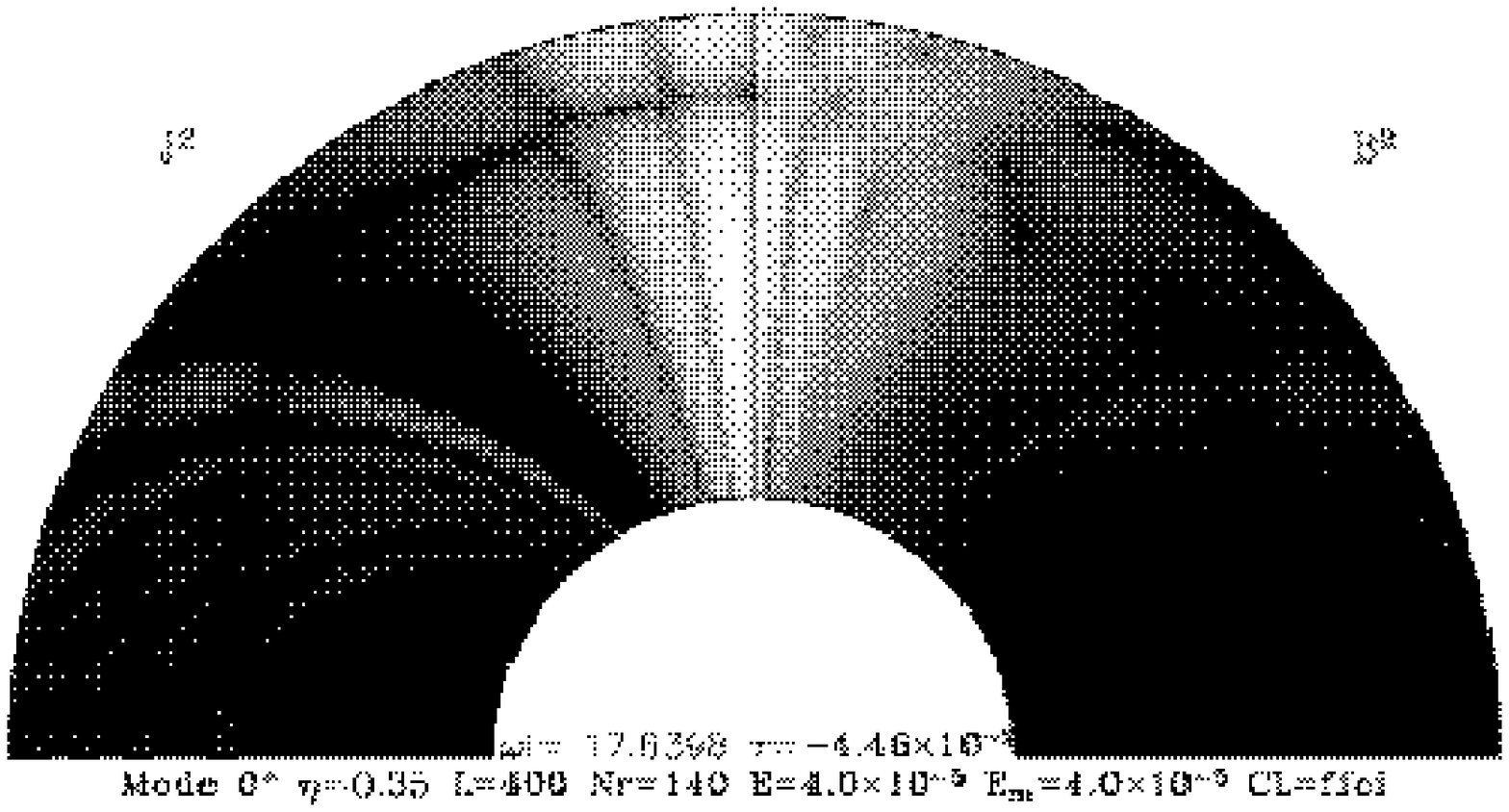} &
\includegraphics[width=8.6cm]{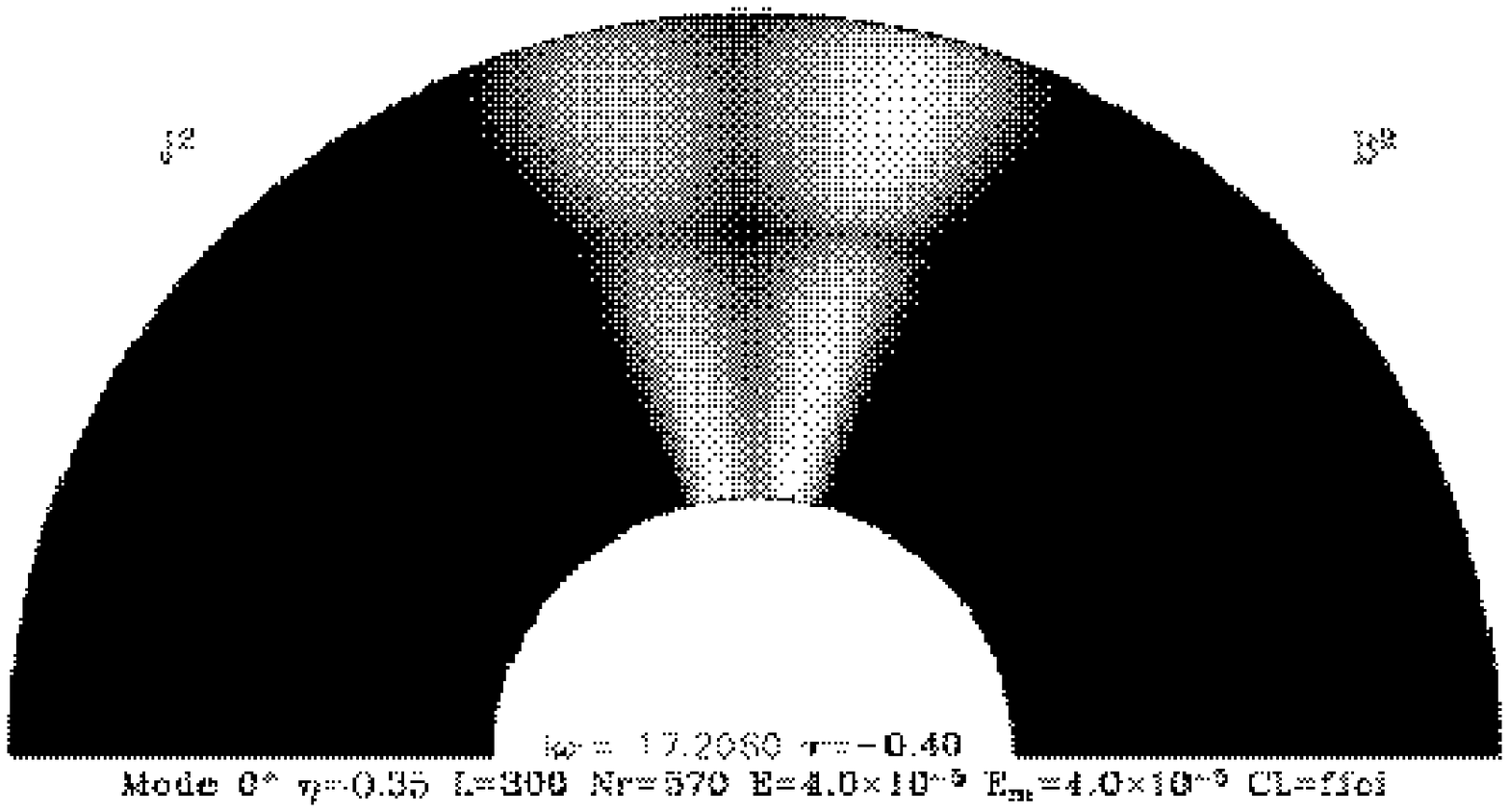} \\
\includegraphics[width=8.6cm]{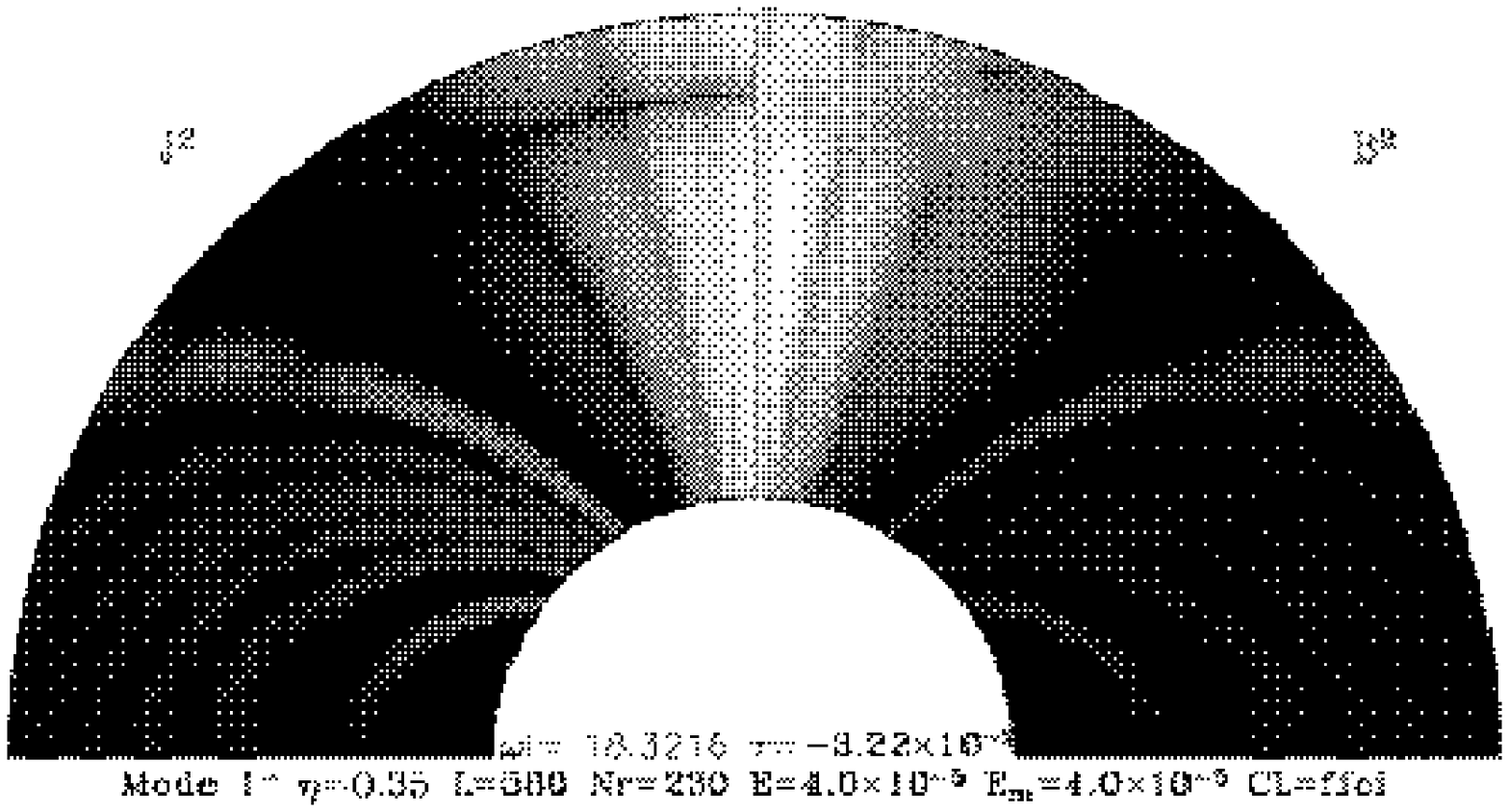} &
\includegraphics[width=8.6cm]{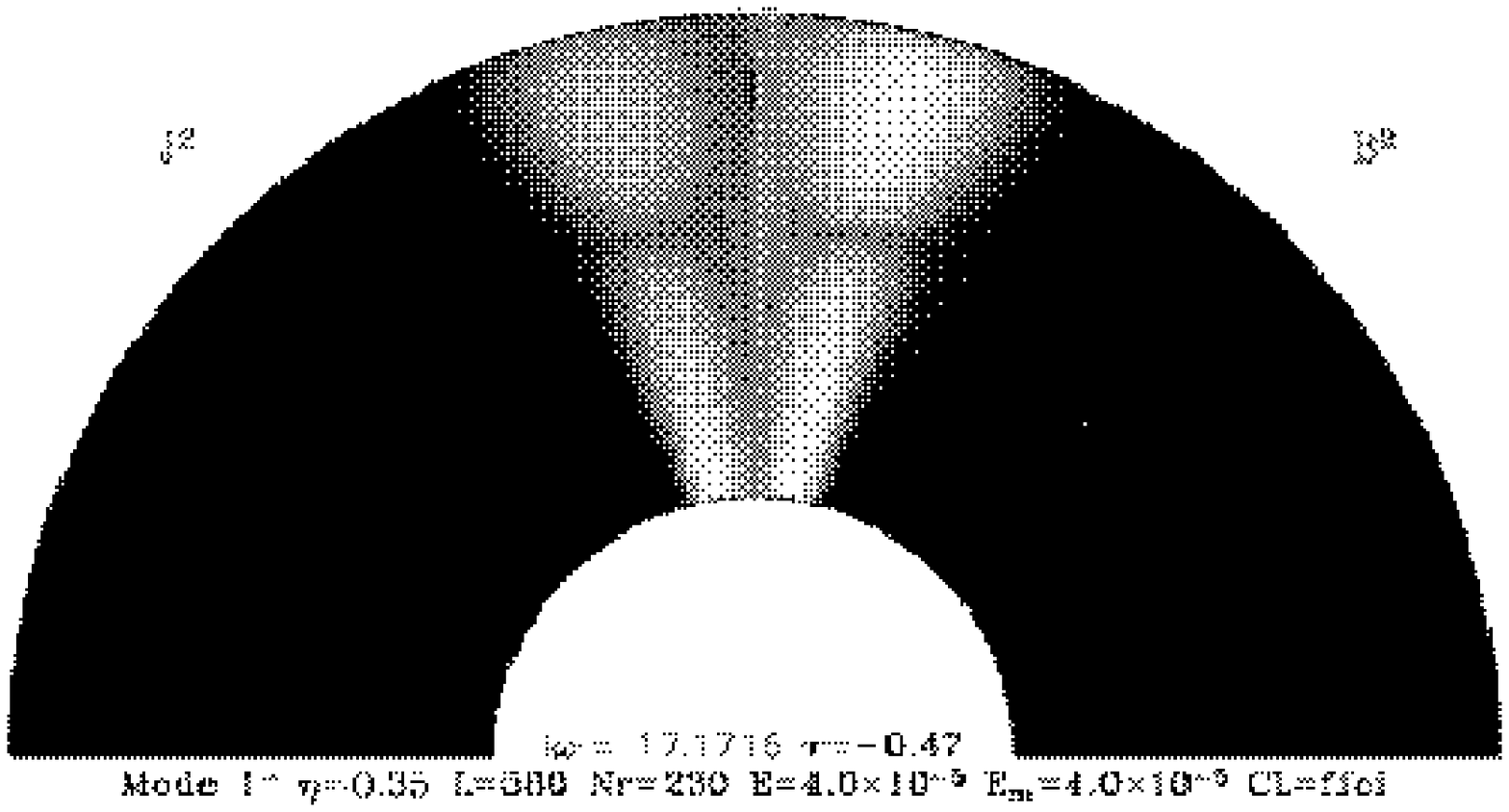}
\end{tabular}
\caption{Comparison between axisymmetric poloidal and toroidal modes and 
non-axisymmetric modes with similar properties. The axisymmetric modes are
above, and the ``corresponding'' non-axisymmetric ones below.}
\label{fig:comparison.modes}
\end{figure*}

\begin{figure*}[htb]
\Large
\begin{tabular}{cc}
\includegraphics[width=8.6cm]{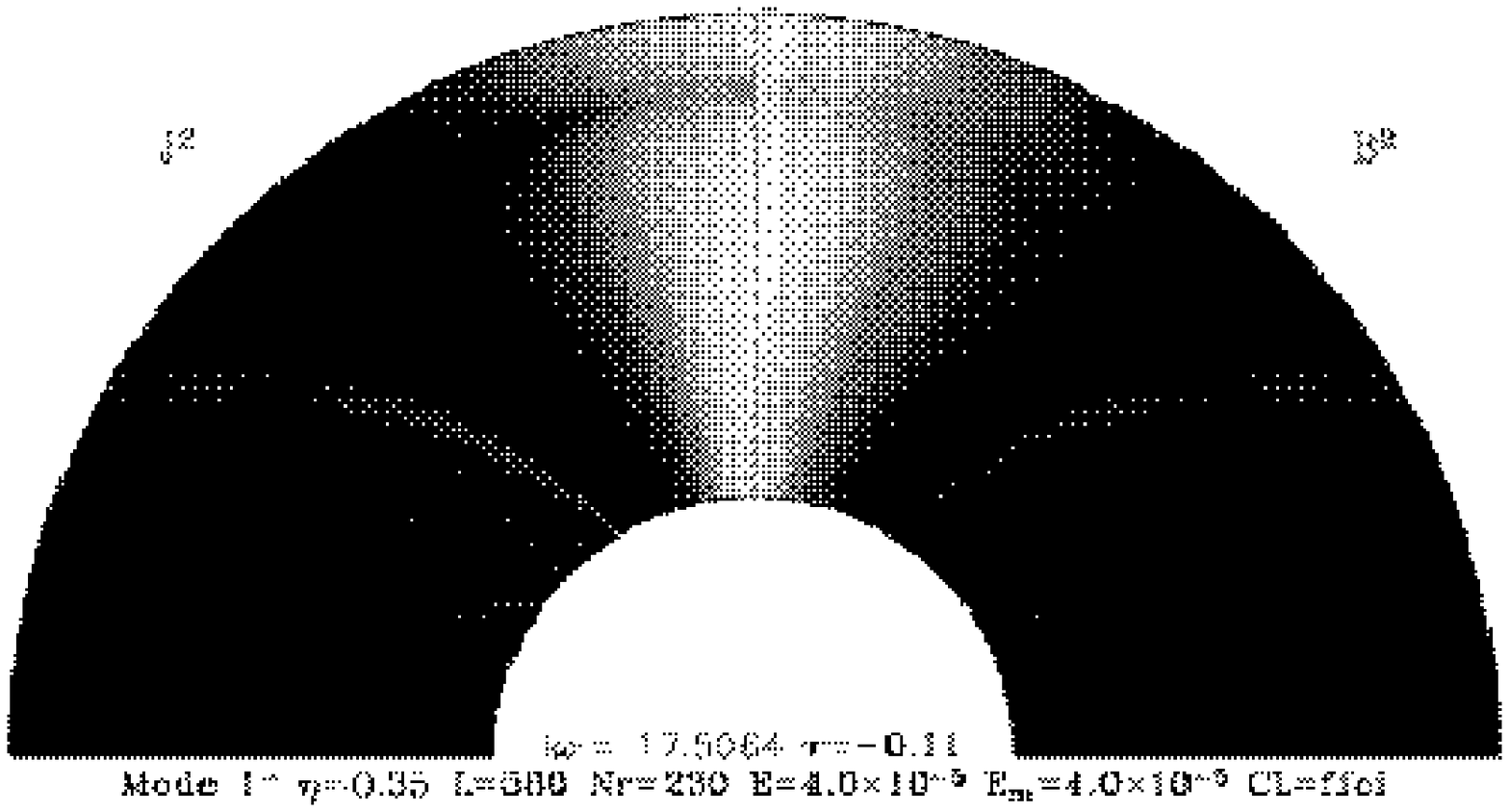} &
\includegraphics[width=8.6cm]{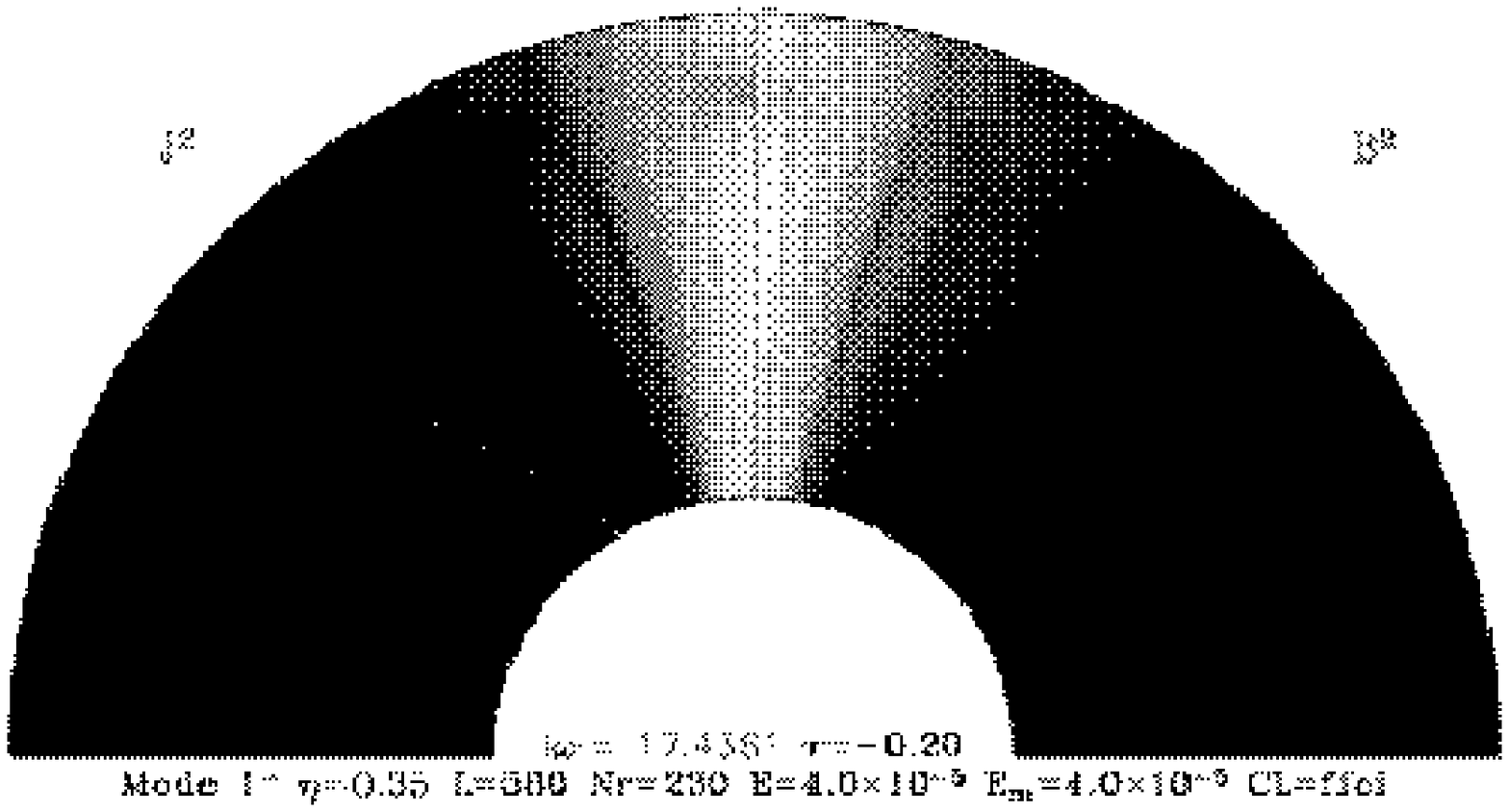} \\
a & b \\
\includegraphics[width=8.6cm]{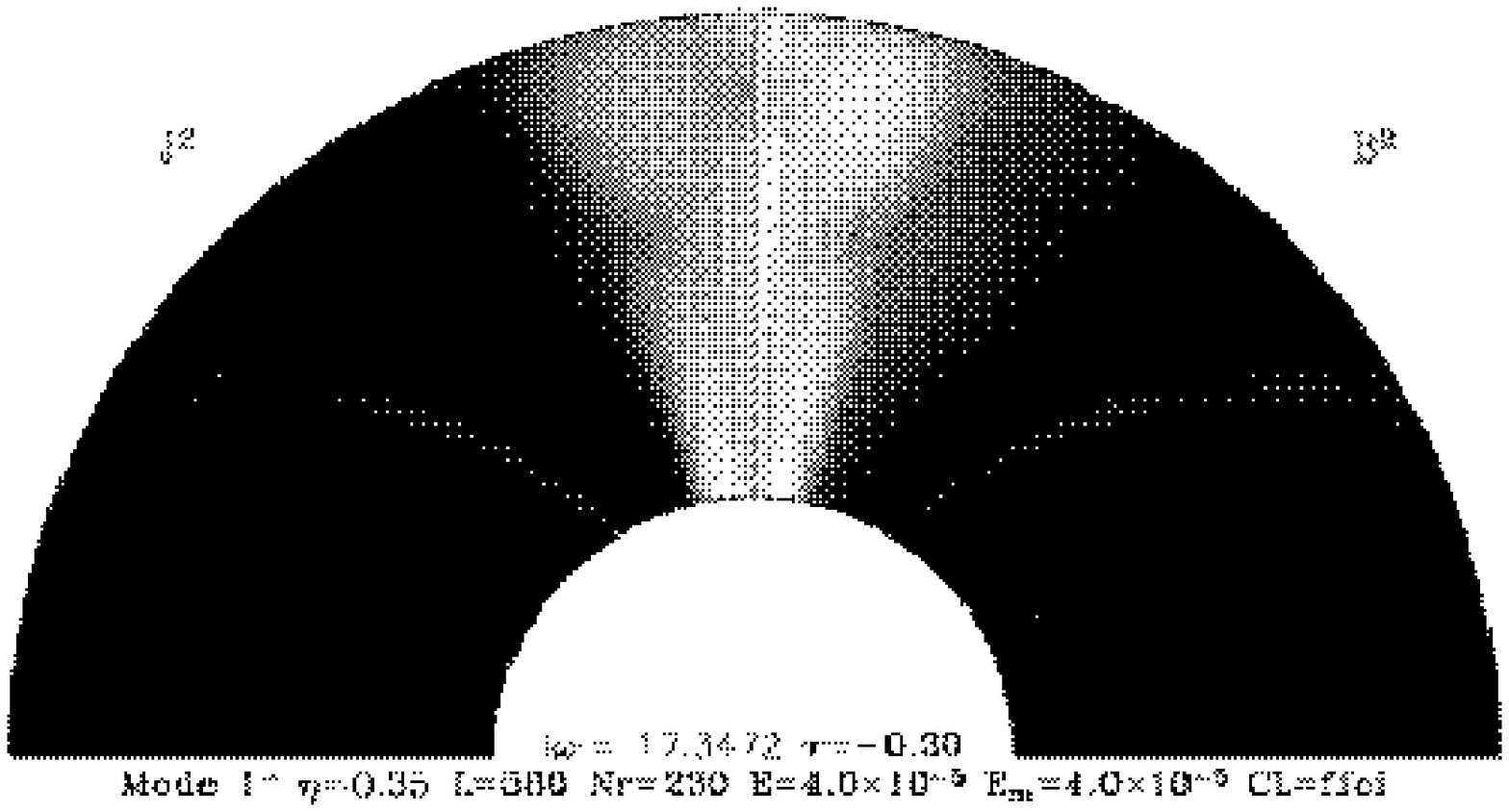} &
\includegraphics[width=8.6cm]{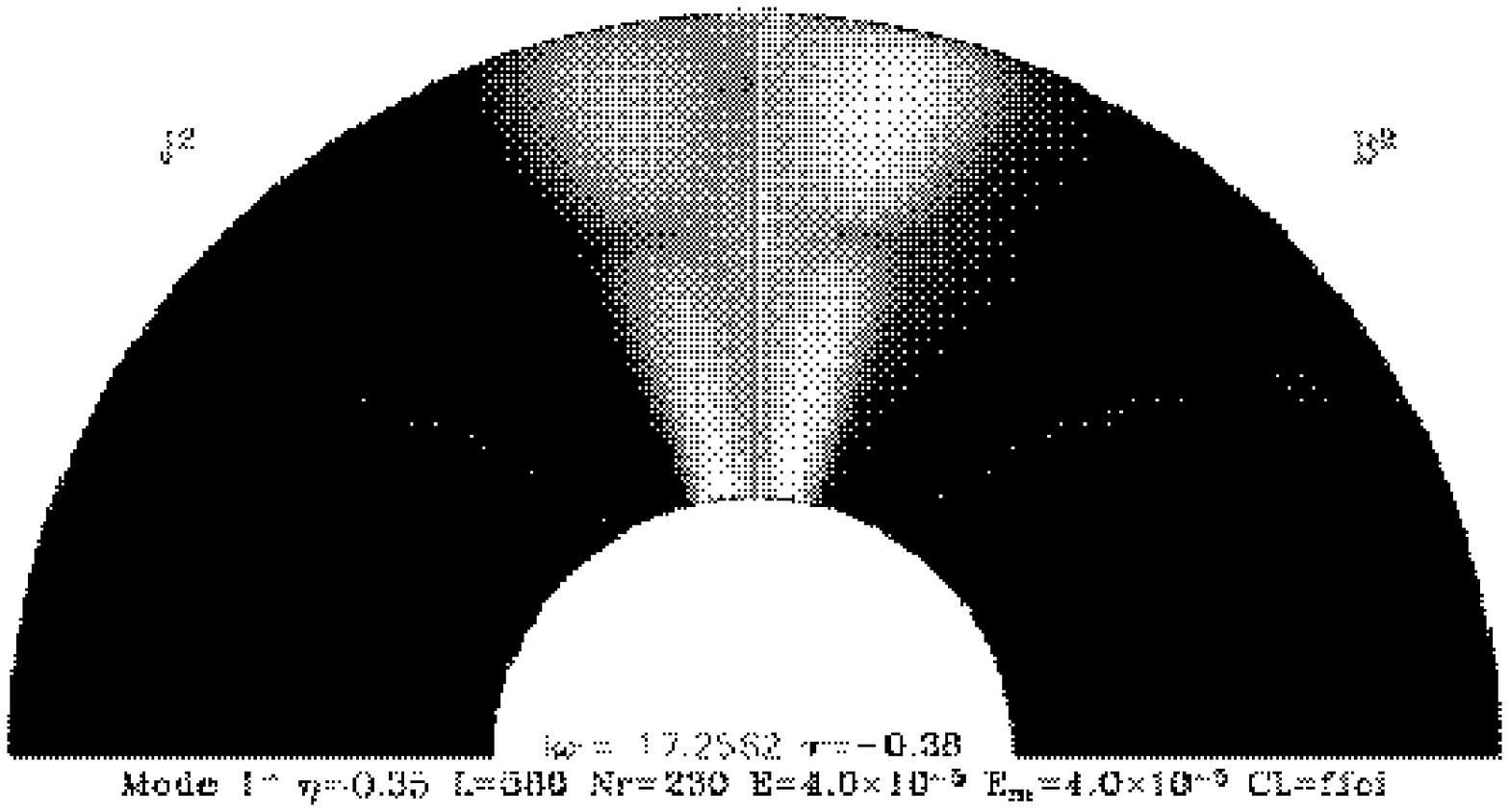} \\
c & d \\
\end{tabular}
\normalsize
\caption{Transition from poloidal to toroidal characteristics in
non-axisymmetric modes.  The diffusivities are $E=\Em=4.10^{-5}$ and $m$ is
equal to $1$.  The letters underneath each diagram correspond to labels on
Fig.~\ref{fig:comparison}.}
\label{fig:transition}
\end{figure*}

It is then interesting to look at the structure of the corresponding modes. 
Not surprisingly, modes along the lower branch have a very similar appearance
to that of poloidal modes.  The radial nodes are in the same positions and the
horizontal nodes  look the same.  As for the modes on the upper branches, it
appears that the least-damped ones are poloidal-like, and the most damped
toroidal-like.  Figure~\ref{fig:comparison.modes} shows a comparison between
axisymmetric poloidal and toroidal modes and non-axisymmetric modes with
similar appearances.  Figure~\ref{fig:transition} shows the transition from
poloidal properties to toroidal ones, when looking at successive modes on the
upper branch.

One way to characterise whether a mode is more poloidal or toroidal is
to calculate the following ratio:
\begin{equation}
\label{eq:ratio}
\chi \! =\frac{H(\ulm)+H(\vlm)+H(\alm)+H(\blm)}{H(\wlm)+H(\clm)},
\end{equation}
where $H(x)$ represents the energy of component $x$.  {\bf $\chi$ is the ratio
of ``poloidal'' energy (both kinetic and magnetic) to ``toroidal'' energy.
Explicitly, it is given by:}
\begin{equation}
\chi \! =\frac{\displaystyle \sum_{\l=\l_{\min}}^{L} \!\! \int_{\eta}^1 \!\!\left[ |\ulm|^2 +
|\alm|^2 + \l(\l+1) \left( |\vlm|^2 + |\blm|^2 \right)\right] r^2 dr }{\displaystyle
\sum_{\l=\l_{\min}}^{L} \int_{\eta}^1  \l(\l+1)\left( |\wlm|^2 + |\clm|^2
\right) r^2 dr},
\end{equation}
When this test is applied to the modes of Fig.~\ref{fig:transition}, we confirm
that ``poloidal'' modes have strong $\alm$,$\blm$,$\ulm$ and $\vlm$ components
whereas ``toroidal'' ones have strong $\wlm$ and $\clm$ components (see
Tab.~\ref{table:poloidal_ratios}).

\begin{table}[h!]
   \begin{center}
   \begin{tabular}{ccc}
      \hline
      \noalign{\smallskip}
      Mode & $ \chi $ \\
      \noalign{\smallskip}
      \hline
      \noalign{\smallskip}
      a & 17.4   \\   
      b & 3.16   \\   
      c & 0.209  \\   
      d & 0.0153 \\   
      \noalign{\smallskip}
      \hline
   \end{tabular}
   \end{center}
   \caption{Energy ratios as defined in Eq.~(\ref{eq:ratio}) for modes shown in
   Fig.~\ref{fig:comparison} and~\ref{fig:transition}.}
   \label{table:poloidal_ratios}
\end{table}

\subsection{Asymptotic numerical behaviour}

\begin{figure}[tb]
\includegraphics[width=8.8cm]{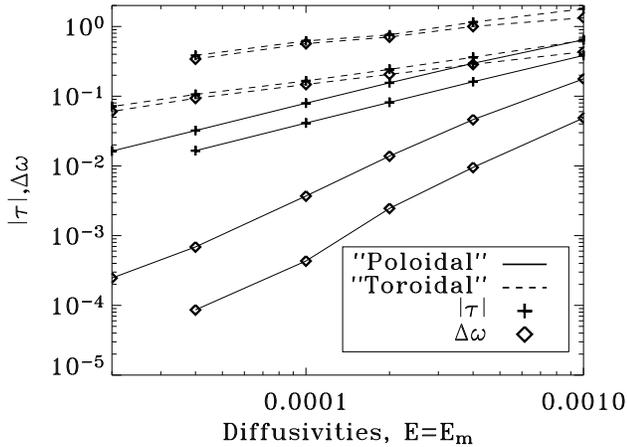}
\caption{Behaviour of $\tau$ and $\Delta \omega$ ($=\omega_{i} - \omega$) for
four non-axisymmetric modes.  The toroidal modes start off with a poloidal
structure, and undergo a metamorphosis into a toroidal appearance.  The
poloidal non-axisymmetric modes approach their inviscid values much faster.}
\label{fig:evolution_m=1}
\end{figure}

We can take a more detailed look at the behaviour of an entire pair of branches
for different values of $E$ and $\Em$ (keeping $E=\Em$), to see whether the
eigenvalues have poloidal or toroidal characteristics. Non-axisymmetric
``poloidal'' eigenvalues appear to obey the following empirical law:
\begin{equation}
\lambda = i\omega^{0} + i\omega^1 E^{\gamma} + \tau^1 E^{\delta},
\end{equation}
where $\gamma \sim 1.5$ and $\delta \sim 1$. This is different from what was
observed in the case of axisymmetric toroidal modes, which show a slower
convergence toward an asymptotic adiabatic value.  In addition each
non-axisymmetric poloidal eigenvalue seems to have its own asymptotic
frequency, as may also {\bf be} the case for axisymmetric poloidal modes.  A
plot of the behaviour of different eigenvalues with respect to diffusivities is
given in Fig.~\ref{fig:evolution_m=1}.   $\omega_i$ was estimated using a
least-square method for the poloidal modes and differs for each one of them.

Non-axisymmetric ``toroidal'' eigenvalues behave roughly like their
axisymmetric counterparts.  All the values on a single branch seem to converge
toward one frequency, the corresponding polar eigenvalue (see
Eq.~(\ref{eq:eigenvalues.polar})).  This value is used in
Fig.~\ref{fig:evolution_m=1} as the asymptotic frequency of the toroidal modes.
The rate at which eigenvalues converge is about the same as in the
axisymmetric case, namely $\lambda_i - \lambda \propto E^{1/2}$.

\begin{figure}[htbp]
\includegraphics[width=8.6cm]{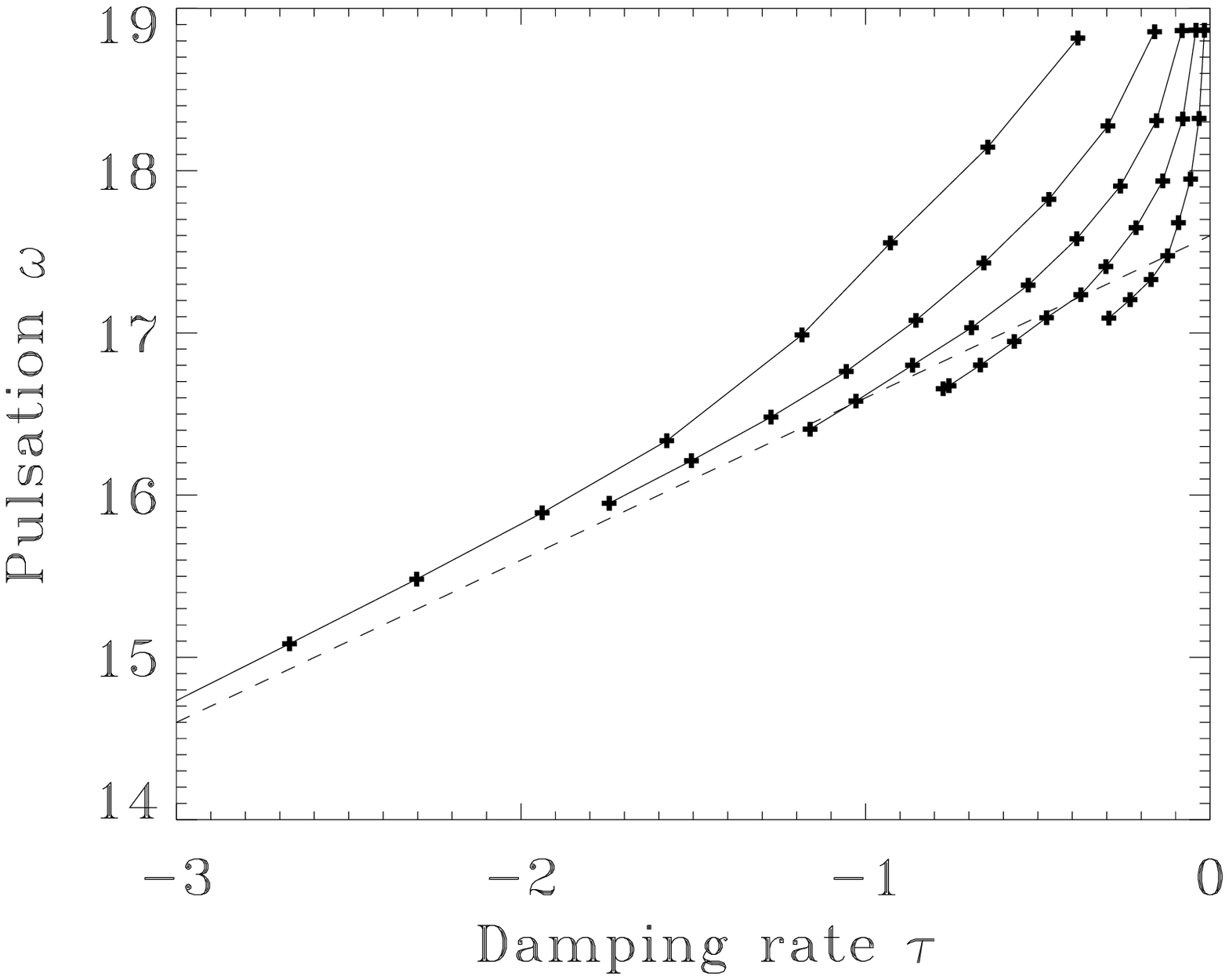}
\includegraphics[width=8.6cm]{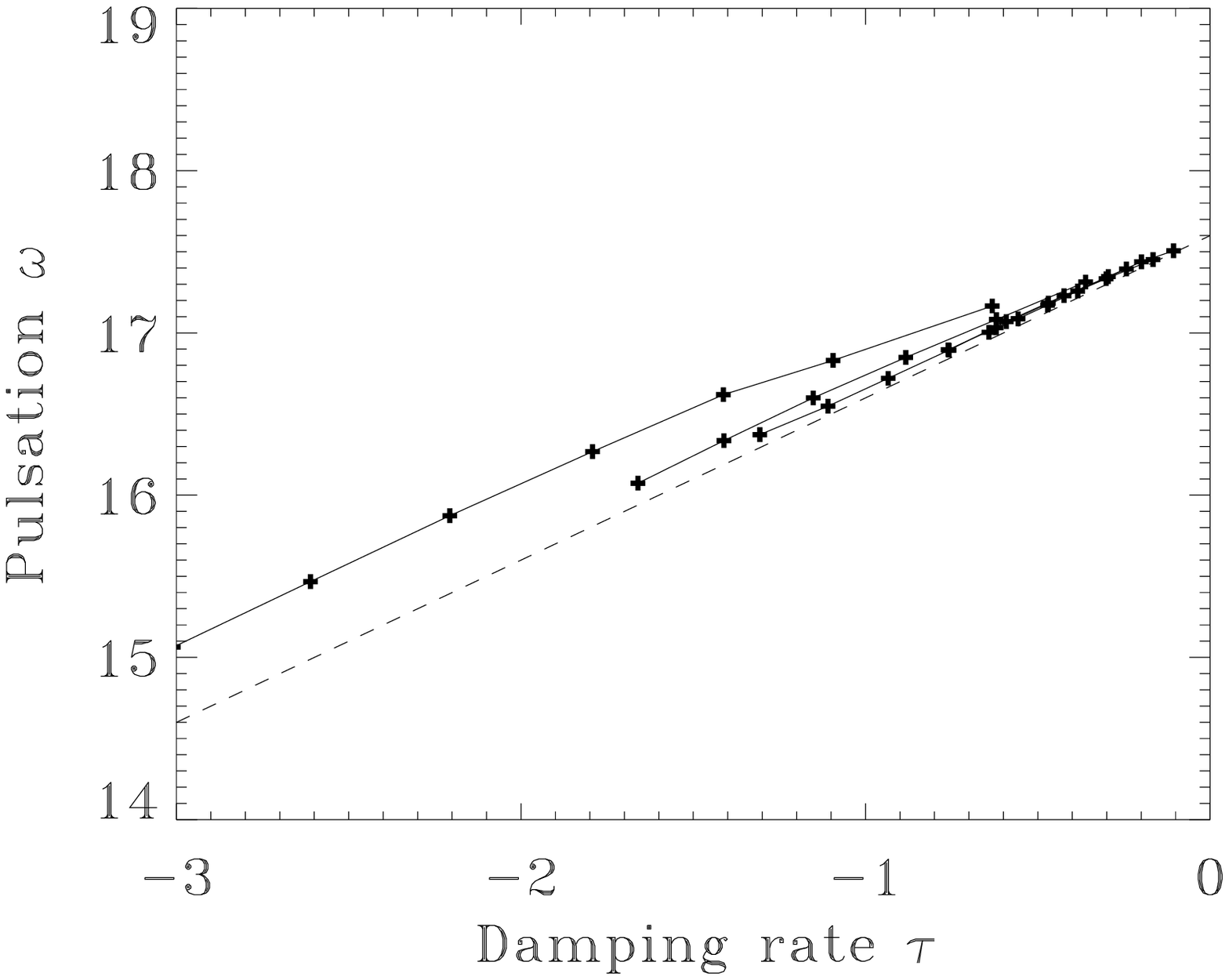}
\caption[]{Behaviour of non-axisymmetric eigenbranches for different values
of the diffusivities. $E(=\Em)$ takes on the values $10^{-3}, 4.10^{-4},
2.10^{-4}, 10^{-4}$ and $4.10^{-5}$.  As diffusivities diminish, both
eigenbranches get closer to the imaginary axis since $|\tau |$ decreases.  The
first figure corresponds to ``poloidal'' eigenvalues and the second one to
``toroidal'' ones.  The classification between the two types of values was done
by hand.  The dotted line represents the line given by Eq.~(\ref{eq:eigenlines})
for $n=1$.}
\label{fig:evolution_branch_m=1}
\end{figure}

It is also interesting to note that the shape of the toroidal eigenbranch
remains more or less the same in the axisymmetric and non-axisymmetric cases
(see Fig.~\ref{fig:evolution_branch_m=1}, {\bf bottom}). This indicates that the
quantization of these modes in the horizontal direction is very similar to
the one of the axisymmetric modes.  However, the spacing between each mode is
not the same (see Fig.~\ref{fig:comparison}).

\begin{figure}[htbp]
\includegraphics[width=8.6cm]{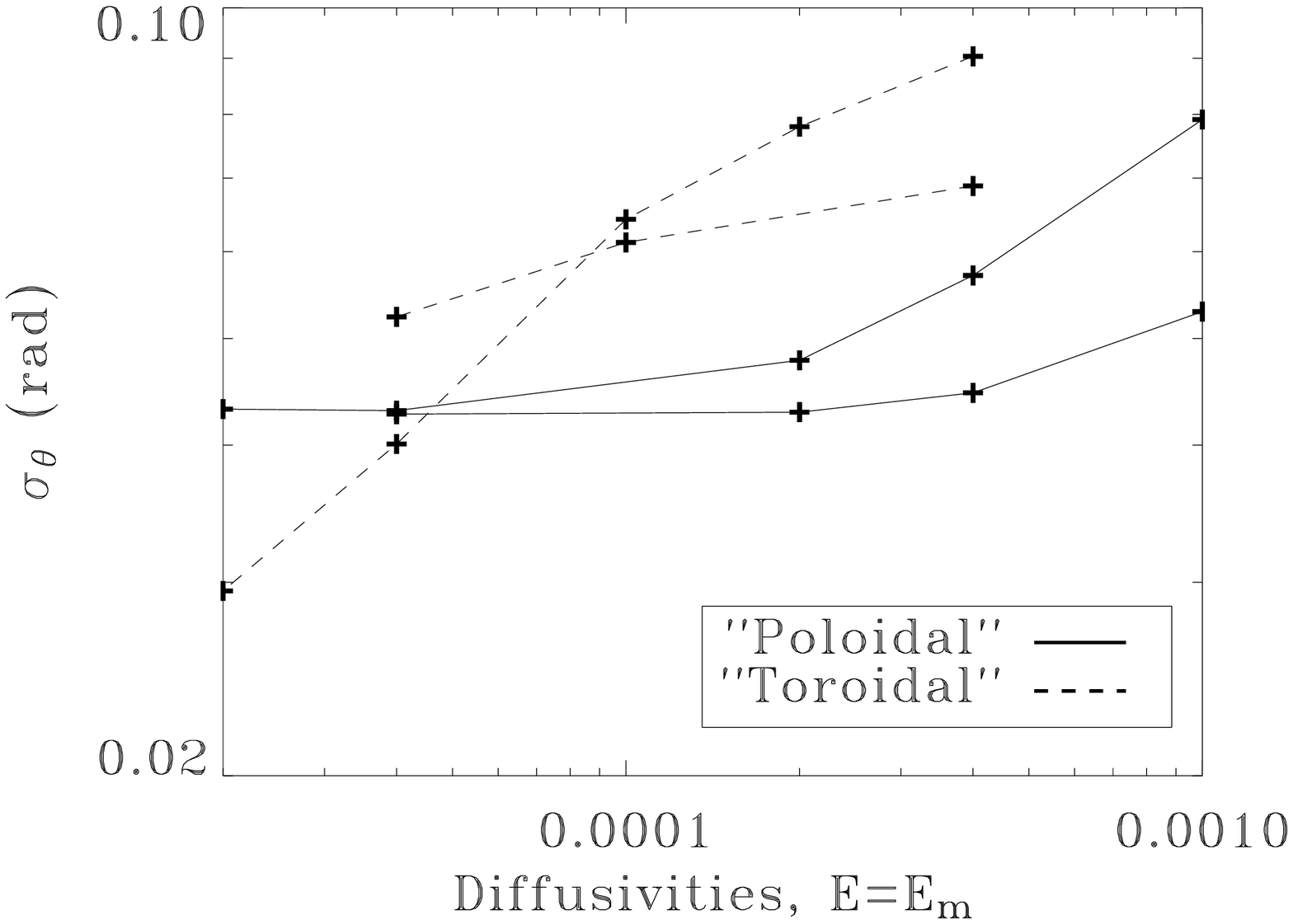}
\caption{Behaviour of non-axisymmetric modes for different values of the
diffusivities.  The ``poloidal'' modes seem to converge toward a
{\bf non-singular} structure.  The ``toroidal'' ones, on the other hand, do not
stop shrinking.  The behaviour of their mean position (not shown), however, is
much more irregular.}
\label{fig:one_plus_p_t}
\end{figure}

By looking at the behaviour of the structure of different modes, it is possible
to observe different characteristics (see Fig.~\ref{fig:one_plus_p_t}). 
Poloidal-like modes seem to converge toward a non-degenerate structure.  In
other words, their widths ($\sigma_{\theta}$) do not approach zero, nor do
their positions ($\overline{\theta}$). Their nodes become more distinct at low
diffusivities.

Toroidal-like modes, on the other hand, exhibit more complex behaviour as the
toroidal character appears only below some given diffusivity. We understand
this behaviour as the consequence of the $E^{1/4}$-scale of toroidal modes. It
is only when $E^{1/4} \ll 1$ that the toroidal character may appear but this
condition is not easy to satisfy with non-axisymmetric modes for which the
lowest diffusivities reached yet are $\sim 10^{-5}$ because of memory storage
requirements.

\section{Discussion}

The main result of this study is that there are two basic behaviours for shear
Alfvén modes, both in the axisymmetric and the non-axisymmetric cases.  The
first one is the poloidal behaviour, for which modes remain regular as
diffusivities approach zero, except for internal shear layers located on
resonant field lines.  The second one is the toroidal behaviour: these modes
become singular as diffusivities approach zero and each branch appears to have
only one eigenfrequency in the adiabatic limit.

{\bf  The toroidal behaviour matches the description of the resistive Alfvén
spectrum given by \citet{Kerner86} in the context of tokamak-like
configurations.   The eigenvalues lie on well-defined curves in the complex
plane which join the endpoints of the ideal Alfvén continua, as is suggested by
Eq.~(\ref{eq:eigenlines}).  The slope of the eigenbranches is $45°$ and the
damping rates of the eigenmodes are proportional to $\E^{1/2}$, which is in
complete agreement with a quadratic variation of the permanent magnetic field
$\theta$-profile near the magnetic axis.

Poloidal modes on the other hand, look like true eigenmodes of the ideal MHD
operator with a finite number of resonances in the Alfvén continuum (resonating
field lines, see paper I). Such features are similar to  those of ``quasimodes"
\cite[see][]{Poedts91,GAA02} which combine a regular and singular behaviour;
however, unlike quasimodes, the damping rate of our modes always seem to vanish
in the ideal limit (e.g. Fig.~\ref{fig:evolution_m=1} or Fig.~5 of Paper~I).

Presently, we understand the origin of the difference between the toroidal
and poloidal behaviour as the presence or absence of a coupling between
magnetic field lines by total pressure. Indeed, neglecting diffusion
terms, magnetic and velocity perturbations verify:

\begin{equation}
\label{eq:poloidal}
\begin{array}{rcl}
\displaystyle \dt{ \vect{\v}} &=& -\grad p_*
                      +(\vect{B} \. \vect{\nabla}) \vect{b}
                      +(\vect{b} \. \vect{\nabla}) \vect{B} \\
\noalign{\smallskip}
\displaystyle \dt{ \vect{b}}  &=&  
                       (\vect{B}  \. \vect{\nabla}) \vect{\v}
                      -(\vect{\v} \. \vect{\nabla}) \vect{B}
\end{array}
\end{equation}
where $p_* = p + \vect{B} \. \vect{b}$ is the perturbed total pressure.  These
equations show that the time evolution of a perturbation depends on the
\textit{global} form of the perturbation only through the pressure gradient
term.  This is clear if one considers a local perturbation, namely an Alfvén
wave packet. Terms like $(\vect{B} \. \vect{\nabla}) \vect{b}$ or $(\vect{B} 
\. \vect{\nabla}) \vect{\v}$ modulate the evolution of the perturbation
according to its derivative along the field lines while terms like $(\vect{b}
\. \vect{\nabla}) \vect{B}$ and $(\vect{\v} \. \vect{\nabla}) \vect{B}$
introduce a dependence with respect to the local value of the perturbation.
Last, the term $\grad p_*$ makes the evolution of the perturbation sensitive to
the pressure perturbation which cannot be restricted to a single field line. In
pure toroidal modes (axisymmetric ones) this term is zero and therefore no
coupling is possible in the ideal MHD case. In non-axisymmetric modes of
toroidal type it is weak and strengthens as one shifts to modes with a strong
poloidal character. Modes studied by \citet{Kerner86} have a weak,
$\O(\Em^{1/2})$, pressure term and are thus similar to our toroidal modes. }

\section{Conclusions}

In this paper we completed the study of shear Alfvén modes of a
spherical shell started in Paper~I. The global picture that emerges is
that both non-axisymmetric and axisymmetric modes can be divided into
two subclasses: the toroidal and poloidal ones. In the limit of small
diffusivities, relevant for astrophysical objects, only poloidal modes
survive since toroidal ones become singular as magnetic shear layers.

However, toroidal modes may play some important role in the excitation process,
especially if rotation is present. Indeed, it is well-known
\cite[e.g.][]{rieu91} that the Coriolis force induces a coupling between
toroidal and poloidal components, even in the axisymmetric case. This coupling
may not strongly influence the frequency of the modes (except for some
rotational splitting) but may represent a channel by which modes lose energy
through dissipative shear layers; in such a case this mechanism may decide
whether a mode is or is not excited by, say, a $\kappa$-mechanism.

For astrophysical applications where the compressibility is an important
aspect, pure Alfvén modes coexist with magneto-acoustic modes. {\bf The
incorporation of these modes will undoubtedly lead to a much more complicated
spectrum of eigenvalues as well as a whole new range of physical phenomena.  In
particular, we can expect new resonances, namely those coming from the
interaction of eigenmodes and the slow acoustic, or cusp, continuum
\cite[see][]{SGH91}. Taking into account compressibility will however not
affect the splitting between pure poloidal and pure toroidal axisymmetric
modes; because of the divergence-less nature of purely toroidal modes, we can
expect that these modes will not be affected by compressibility and remain
unchanged. All the complications will concentrate on the poloidal set of
eigenmodes. In the case of non-axisymmetric modes where the poloidal and
toroidal characters are mixed, we may expect that only modes with a strong
poloidal shape will suffer important changes. However, the importance of these
changes, which incorporate the resonances with the cusp continuum, as well as
its influence on the shape of the spectrum, can hardly be anticipated and will
be investigated in subsequent work. }

\begin{acknowledgements}
{\bf The authors wish to thank the referee for valuable suggestions and comments,
which helped to improve this article.}

Some of the numerical calculations
were carried out on the NEC SX5 of the ``Institut du Développement et des
Ressources en Informatique Scientifique'' (IDRIS), which is gratefully
acknowledged.
\end{acknowledgements}

\bibliographystyle{aa}
\bibliography{0539bib}

\begin{thebibliography}{20}
\expandafter\ifx\csname natexlab\endcsname\relax\def\natexlab#1{#1}\fi

\bibitem[{{Abramowitz} \& {Stegun}(1972)}]{Abramowitz}
{Abramowitz}, M. \& {Stegun}, I.~A., eds. 1972, {Handbook of Mathematical
  Functions} (New York: Dover)

\bibitem[{{Alfvén}(1945)}]{Alfven45}
{Alfvén}, H. 1945, MNRAS, 105, 3

\bibitem[{{Balmforth} {et~al.}(2001){Balmforth}, {Cunha}, {Dolez}, {Gough}, \&
  {Vauclair}}]{excitation}
{Balmforth}, N.~J., {Cunha}, M.~S., {Dolez}, N., {Gough}, D.~O., \& {Vauclair},
  S. 2001, MNRAS, 323, 362

\bibitem[{{Bigot} \& {Dziembowski}(2002)}]{ObliqueModelRevisited}
{Bigot}, L. \& {Dziembowski}, W.~A. 2002, A\&A, 391, 235

\bibitem[{{B}iront {et~al.}(1982){B}iront, {G}oossens, {C}ousens, \&
  {M}estel}]{biront}
{B}iront, D., {G}oossens, M., {C}ousens, A., \& {M}estel, L. 1982, MNRAS, 201,
  619

\bibitem[{{C}ampbell \& Papaloizou(1986)}]{campbell}
{C}ampbell, C.~G. \& Papaloizou, J. C.~B. 1986, MNRAS, 220, 577

\bibitem[{{Chandrasekhar}(1961)}]{Chandrasekhar}
{Chandrasekhar}, S. 1961, {Hydrodynamic and Hydromagnetic Stability} (New York:
  Dover), pp. 155-157

\bibitem[{{Dziembowski} \& {Goode}(1996)}]{Dziembowski}
{Dziembowski}, W.~A. \& {Goode}, P.~R. 1996, ApJ, 458, 338

\bibitem[{{Goossens} {et~al.}(2002){Goossens}, {Andries}, \&
  {Aschwanden}}]{GAA02}
{Goossens}, M., {Andries}, J., \& {Aschwanden}, M.~J. 2002, A\&A, 394, L39

\bibitem[{{Kato} \& {Watanabe}(1958)}]{Kato58}
{Kato}, Y. \& {Watanabe}, T. 1958, JGR, 63, 741

\bibitem[{{Kerner} {et~al.}(1986){Kerner}, {Lerbinger}, \& {Riedel}}]{Kerner86}
{Kerner}, W., {Lerbinger}, K., \& {Riedel}, K. 1986, Phys. Fluids, 29, 2975

\bibitem[{{Kurtz}(1978)}]{1978IBVS}
{Kurtz}, D.~W. 1978, Informational Bulletin on Variable Stars, 1436, 1

\bibitem[{{Lou}(1995)}]{lou95}
{Lou}, Y. 1995, \mnras, 275, L11

\bibitem[{{Poedts} \& {Kerner}(1991)}]{Poedts91}
{Poedts}, S. \& {Kerner}, W. 1991, Phys. Rev. Letters, 66, 2871

\bibitem[{{Poth\'erat} {et~al.}(2002){Poth\'erat}, {Sommeria}, \&
  {Moreau}}]{CoucheHartmann}
{Poth\'erat}, A., {Sommeria}, J., \& {Moreau}, R. 2002, Phys. Fluids, 14, 403

\bibitem[{Rieutord(1991)}]{rieu91}
Rieutord, M. 1991, Geophys. Astrophys. Fluid Dyn., 59, 185

\bibitem[{{Rincon} \& {Rieutord}(2003)}]{rincon}
{Rincon}, F. \& {Rieutord}, M. 2003, A\&A, 398, 663

\bibitem[{{R}oberts \& {S}oward(1983)}]{roberts}
{R}oberts, P.~H. \& {S}oward, A.~M. 1983, MNRAS, 205, 1171

\bibitem[{{Sakurai} {et~al.}(1991){Sakurai}, {Goossens}, \& {Hollweg}}]{SGH91}
{Sakurai}, T., {Goossens}, M., \& {Hollweg}, J.~V. 1991, Solar Phys., 133, 227

\bibitem[{{Stevenson}(1983)}]{stevenson83}
{Stevenson}, D.~J. 1983, Reports of Progress in Physics, 46, 555

\end{thebibliography}
\addcontentsline{toc}{chapter}{References}
\normalsize

\appendix
\section{$\theta$-dependence of asymptotic axisymmetric toroidal modes}
\label{appendix:latitudinal}

We begin with the toroidal equations (\ref{eq:toroidal}), and express
them explicitly in ``semi-dipolar'' coordinates $(r,\nu,\varphi)$.  The
coordinate $\nu$ is equal to $\sin \theta / \sqrt{r}$, and remains constant
along field lines.  Choosing $\nu$ as a coordinate instead of $\theta$
simplifies the expression of terms like $(\vect{B} \cdot \nabla)\vect{b}$.  We
obtain the following system:

\begin{equation}
\begin{array}{lcl}
  \lambda b &=& \displaystyle \sqrt{1- r \nu^2}
     \left[ \frac{1}{r^3} \dpart{\v}{r} - \frac{3\v}{2 r^4} \right]
    + \Em \lapl' b,  \\
  \noalign{\smallskip}
  \lambda \v &=& \displaystyle \sqrt{1- r \nu^2}
     \left[ \frac{1}{r^3} \dpart{b}{r} + \frac{3b}{2 r^4} \right]
    + E  \lapl' \v,
\end{array}
\end{equation}
where $\lapl'$ takes on the following expression :

$$ 
\begin{array}{lcl}
\lapl' b &=& \displaystyle \ddpart{b}{r} 
            - \frac{\nu}{r} \frac{\partial^2 b}{\partial \nu \partial r}
            - \frac{3\nu^2}{4r^2} \ddpart{b}{\nu}
            - \frac{9\nu}{4r^2} \dpart{b}{\nu} \\
\noalign{\smallskip}
        & & \displaystyle
            + \frac{2}{r} \dpart{b}{r}
            +\frac{1}{r^3} \left[ \ddpart{b}{\nu} 
                     + \frac{1}{\nu} \dpart{b}{\nu}
                     - \frac{b}{\nu^2} \right].
\end{array}
$$

{\bf In order to analyse} the asymptotic limit of small diffusivities, we
suppose $E$ and $\Em$ are of the form $K\E$ and $\Km \E$ respectively, where
$K$ and $\Km$ are two constants, and $\E$ a free parameter that approaches
zero.  We then rescale $\nu$ to $\hat{\nu} = \E^{-1/4} \nu$, in accordance with
our numerical results, and neglect terms smaller than $\O (\E^{1/2})$.  This
leads to the following system:
\begin{equation}
\label{eq:toroidal.asymptotic}
\begin{array}{lcl}
  \lambda b \!\! &=& \displaystyle \!\!
     \left( 1- \frac{1}{2} \E^{1/2} r \hat{\nu}^2 \right) \!\!
     \left[ \frac{1}{r^3} \dpart{\v}{r} - \frac{3\v}{2 r^4} \right] 
     \!\! + \! \frac{\E^{1/2}\Km}{r^3} \Theta [b], \\
  \noalign{\smallskip}
  \lambda \v \!\! &=& \displaystyle \!\!
     \left( 1- \frac{1}{2} \E^{1/2} r \hat{\nu}^2 \right) \!\!
     \left[ \frac{1}{r^3} \dpart{b}{r} + \frac{3b}{2 r^4} \right] 
     \!\! + \! \frac{\E^{1/2}K}{r^3} \Theta [\v],
\end{array}
\end{equation}
where the differential operator $\Theta$ is given by:
$$
\Theta [b] =  \ddpart{b}{\hat{\nu}} 
               + \frac{1}{\hat{\nu}} \dpart{b}{\hat{\nu}}
               - \frac{b}{\hat{\nu}^2} .
$$             
Following the form of Eq.~(\ref{eq:empirical.eigenbranch.bis}), we develop the
solutions in the following manner:
\begin{equation}
\begin{array}{lclcl}
  \lambda &=& \lambda^0 &+& \E^{1/2} \lambda^{1}, \\
  b       &=& b^{0}  &+& \E^{1/2}b^{1}, \\
  \v      &=& \v^{0} &+& \E^{1/2}\v^{1}.
\end{array}
\end{equation}
At zeroth order, we obtain the equations that give the polar eigenspectrum (see
Eq.~(\ref{eq:toroidal.ad}) for $\cos \theta = 1$).  We conclude that the
zeroth order solutions to Eq.~(\ref{eq:toroidal.asymptotic}) are of the form:
\begin{equation}
\begin{array}{lcl}
b^{0}(r,\hat{\nu}) &=& b^0_n(r)f(\hat{\nu}), \\
\v^{0}(r,\hat{\nu}) &=& \v^0_n(r)f(\hat{\nu}),\\
\lambda^{0} &=& \lambda^0_n,
\end{array}
\end{equation}
where $(b^0_n,\v^0_n)$ is given by Eq.~(\ref{eq:solution.polar}), and
$\lambda^0_n$ is the corresponding eigenvalue.  $\lambda^0_n$ gives us
the position of the different eigenbranches (as already seen in Sect.
\ref{section:polar.eigenspectrum}).  Finding a quantization on the function
$f$ would enable us to explain the different eigenvalues on a given branch. 
This requires looking at the first order of Eq.~(\ref{eq:toroidal.asymptotic})
(in which we substitute results from the zeroth order):
\begin{equation}
\label{eq:first.order}
\begin{array}{lcl}
\displaystyle \lambda^0_n b^1 - 
\frac{1}{r^3} \dpart{\v^1}{r} + \frac{3\v^1}{2 r^4} &=&
-\lambda^1_n b_n^0 f \\
\noalign{\smallskip}
& & \displaystyle 
   -\frac{\lambda^0_n r\hat{\nu}^2 b^0_n f}{2}  
   +\frac{b^0_n \Km \Theta [f]}{r^3}, \\
\noalign{\smallskip}
\displaystyle \lambda^0_n v^1 -
\frac{1}{r^3} \dpart{b^1}{r} - \frac{3b^1}{2 r^4} &=&
-\lambda^1_n v_n^0 f \\
\noalign{\smallskip}
& & \displaystyle 
   -\frac{\lambda^0_n r\hat{\nu}^2 v^0_n f}{2}  
   +\frac{v^0_n K \Theta [f]}{r^3}, \\
\noalign{\smallskip}
b^1(r=1,\hat{\nu}) = 0, \\
\noalign{\smallskip}
\v^1(r=\eta,\hat{\nu}) = 0.  \\
\end{array}
\end{equation}
{\bf (We now use the notation $\lambda_n^1$ for $\lambda^1$).}
This equation is of the form $\mathcal{L}_0 \Psi_1 = \mathcal{L}_1 \Psi_0$,
where $\mathcal{L}_0$, $\mathcal{L}_1$ are respectively zeroth and first order
linear operators, and $\Psi_0$, $\Psi_1$ are zeroth and first order solutions.  It is
solvable only if its right-hand side is orthogonal to the kernel of the adjoint
of the operator $\mathcal{L}_0$, with respect to the following dot product:
\begin{equation}
\left< \left[ \begin{array}{l} u \\ w \end{array} \right], 
       \left[ \begin{array}{l} x \\ y \end{array} \right] \right>
= \int_{\eta}^1 \! \frac{u^*(r)x(r) + w^*(r) y(r)}{\| \vect{B}(r,0) \|} \, dr,
\end{equation}
where $u^*$ is the complex conjugate of $u$, and $\| \vect{B}(r,0) \| = 1/r^3$.
The adjoint operator $\mathcal{L}_0^\dag$ is deduced from $\mathcal{L}_0$ by
using integration by parts:
\begin{equation}
\mathcal{L}_0^{\dag} \left[ \begin{array}{l} u \\ w \end{array} \right] =
\left[
\begin{array}{l}
\displaystyle (\lambda^0_n)^* u + \frac{1}{r^3}\dpart{w}{r}-\frac{3w}{2r^4} \\
\noalign{\smallskip}
\displaystyle (\lambda^0_n)^* w + \frac{1}{r^3}\dpart{u}{r}+\frac{3u}{2r^4}
\end{array}
\right].
\end{equation}
The adjoint boundary conditions are chosen so as to cancel out terms like
$w^*(1)b(1) - w^*(\eta)b(\eta)$, which come from the integration by parts:
\begin{equation}
  u(r=1) = 0, \qquad w(r=\eta) = 0.
\end{equation}
The zeroth order eigenvalue $\lambda_n^0$ is purely imaginary.  As a result the
operator $\mathcal{L}_0$ is antisymmetric: $\mathcal{L}_0^\dag =
-\mathcal{L}_0$.  The boundary conditions being the same for both operators, 
the kernel of $\mathcal{L}_0^\dag$ is the same as that of $\mathcal{L}_0$:
\begin{equation}
u = b_n^0, \qquad w = \v_n^0.
\end{equation}

By calculating the dot product between $(b_n^0,\v_n^0)$ and the terms of
Eq.~(\ref{eq:first.order}) we obtain the following equation:
\begin{equation}
\label{eq:latitudinal}
-\lambda^1_n \mathcal{C}_1 f
-\frac{\lambda_n^0 \hat{\nu}^2 \mathcal{C}_2}{2}f
+\mathcal{C}_3 \Theta [f] = 0
\end{equation}
where:
\begin{equation}
\label{eq:constants}
\begin{array}{lcl}
\mathcal{C}_1 &=& 
\left< \left[ \begin{array}{l} u     \\ w     \end{array} \right], 
       \left[ \begin{array}{l} b^0_n \\ v^0_n \end{array} \right] \right> \\
\noalign{\smallskip}
&=& \displaystyle
 \int_{\eta}^1 \! r^3 \left( |b^0_n|^2 + |v^0_n|^2 \right) \, dr, \\
\noalign{\smallskip}
\mathcal{C}_2 &=& 
\left< \left[ \begin{array}{l} u     \\ w     \end{array} \right], r
       \left[ \begin{array}{l} b^0_n \\ v^0_n \end{array} \right] \right> \\
\noalign{\smallskip}
&=& \displaystyle
 \int_{\eta}^1 \! r^4 \left( |b^0_n|^2 + |v^0_n|^2 \right) \, dr, \\
\noalign{\smallskip}
\mathcal{C}_3 &=& 
\displaystyle
\left< \left[ \begin{array}{l} u     \\ w     \end{array} \right], \frac{1}{r^3}
\left[ \begin{array}{l} \Km b^0_n \\ K v^0_n \end{array} \right] \right> \\
\noalign{\smallskip}
&=& \displaystyle
 \int_{\eta}^1 \! \Km |b^0_n|^2 + K |v^0_n|^2  \, dr.
\end{array}
\end{equation}
Equation~(\ref{eq:latitudinal}) gives the quantization along each eigenbranch.

\section{Simplified analytical model}
\label{appendix:model}

The model presented here justifies the use of effective boundary conditions
throughout the paper for the adiabatic calculations.  In this model, we use
Cartesian coordinates $(x,y,z)$, and the fluid domain is enclosed between two
infinite horizontal planes located at $z=0$ and $z=1$. The permanent magnetic
field is constant and in the vertical direction $(\vect{B}=B\ez)$.  The
perturbations take on the form $ \vect{b} = b(z)\,\ey $ and $ \vect{\v} =
\v(z)\ey $, in order to mimic the toroidal direction $\ephi$.  Equation
(\ref{eq:system}) becomes:
\begin{equation}
\label{eq:simple}
\begin{array}{lcl}
     \lambda b &=& \displaystyle \frac {\d \v} {\d z} + \Em \frac {\d^2 b}{\d z^2}, \\
     \noalign{\smallskip}
     \lambda \v &=& \displaystyle \frac {\d b} {\d z} + E \frac {\d^2 \v}{\d z^2}.
\end{array}
\end{equation}
The corresponding boundary conditions are:
\begin{equation}
\begin{array}{lll}
    \displaystyle \frac{\d \v}{\d z}(0) = 0, & &
    \displaystyle \Em \frac{\d b}{\d z}(0) = -\v(0), \\
    \noalign{\smallskip}
    \displaystyle \frac{\d \v}{\d z}(1) = 0, & & b(1)=0.
\end{array}
\end{equation}
Since the two conditions on the derivative of $\v$ are stress free conditions,
they lack physical meaning when both diffusivities vanish.  Therefore, we can
expect them to disappear in that situation. The solutions to
Eq.~(\ref{eq:simple}) are:
\begin{equation}
\label{eq:solutions}
\begin{array}{lcl}
    \v &=& k_1^{-1}k_2 \sin k_2 \left(\e^{ik_1z}+\e^{ik_1(2-z)} \right) \\
    \noalign{\smallskip}
       & & -i\left( 1-\e^{2ik_1} \right) \cos (k_2(z-1)), \\
    \noalign{\smallskip}
    b  &=& k_2 \sin k_2 \left( \zeta^- k_1^{-1} + i \xi \right)
    \left( \e^{ik_1z}-\e^{ik_1(2-z)} \right) \\
    \noalign{\smallskip}
       & & + \left( 1 - \e^{2ik_1} \right) \left( \zeta^- + 
           i \xi k_2 \right)\sin (k_2(z-1)),
\end{array}
\end{equation}
where\footnote{When taking the square root of a complex number $z$, we choose
the root such that $\arg(\sqrt{z}) \in ]-\pi/2,\pi/2]$}:
\begin{equation}
\begin{array}{lcl}
    \zeta^+ &=& \sqrt {1 + \left( \sqrt{E} + \sqrt{\Em} \right)^2 \lambda}, \\
    \noalign{\smallskip}
    \zeta^- &=& \sqrt {1 + \left( \sqrt{E} - \sqrt{\Em} \right)^2 \lambda}, \\
    \noalign{\smallskip}
    \xi &=& \sqrt{E\Em} - E, \\
    \noalign{\smallskip}
    k_1 &=& \displaystyle \frac{i(\zeta^- + \zeta^+)}{2\sqrt{E\Em}} 
            \sim \frac{i}{\sqrt{E\Em}}, \\
    \noalign{\smallskip}
    k_2 &=& \displaystyle \frac{i(\zeta^- - \zeta^+)}{2\sqrt{E\Em}}
            \sim -i \lambda.
\end{array}
\end{equation}
{\bf The quantities $k_1$, $k_2$, $-k_1$,and $-k_2$ are solutions of the
dispersion relation of Alfvén waves in a uniform field \citep{Chandrasekhar}:
\begin{equation}
  -(\lambda + E k^2)(\lambda + \Em k^2) = k^2
\end{equation}
where the non-dimensional Alfvén velocity is $1$ and the angle between the wave
vector $\vect{k}$ and the permanent magnetic field is $0$.}

The eigenvalue $\lambda$ must satisfy the following relation:
\begin{equation}
\begin{array}{l}
  \left\{ -\zeta^- \lambda +  i[(\zeta^-)^2 + \xi \lambda]k_1 \right\} \sin k_1 \cos k_2 \\
  \noalign{\smallskip}
  \quad = \left\{ -\zeta^- \lambda + i[(\zeta^-)^2 + \xi \lambda]k_2 \right\} \sin
  k_2 \cos k_1.
\end{array}
\end{equation}

If the ratio between $E$ and $\Em$ remains constant, we obtain the following
formula for $\lambda$:
\begin{equation}
\lambda = i \Omega_n - \left( \frac{E+\Em}{2} \right) \Omega_n^2 + \O (E^2),
\label{eq:spectre}
\end{equation}
where $ \Omega_n = \left(n + \frac{1}{2} \right) \pi$.

When $E$ and $\Em$ approach zero, the solutions given in
Eq.~(\ref{eq:solutions}) obey the following effective boundary conditions:
\begin{equation}
\label{eq:effective.conditions}
        \v(0) = 0 \qquad b(1) = 0, 
\end{equation}
and their gradients admit a discontinuity on the lower boundary.  It is
interesting to note that the stress-free condition has disappeared on the lower
boundary, as was expected. On the upper boundary, it simply became redundant
with the condition $b(1)=0$.

When comparing the eigenvalues given in Eq.~(\ref{eq:spectre}) to the ones
given in Eq.~(\ref{eq:tor.eigenvalues}), we notice that the term
$\O(\E^{1/2})$ is lacking in the first formula.  This is because in the
simplified model, eigenmodes are invariant in the $\ex$ direction, thus 
removing any quantization in that direction.

\end{document}